\documentclass[10pt,twocolumn]{IEEEtran}
\ifCLASSOPTIONcompsoc
  \usepackage[nocompress]{cite}
\else
  \usepackage{cite}
\fi

\usepackage{rotating}
\usepackage{tablefootnote}
\usepackage[flushleft]{threeparttable}
\usepackage[utf8]{inputenc} % allow utf-8 input
\usepackage[T1]{fontenc}    % use 8-bit T1 fonts
\usepackage{url}            % simple URL typesetting
\usepackage{array}
\usepackage{multirow}
\usepackage{longtable}
\usepackage[table,xcdraw]{xcolor}
\usepackage{amssymb,amsmath,amsthm,enumitem}
\usepackage{subcaption}
\usepackage{xcolor}
\usepackage{scalerel}
\usepackage{comment}
\usepackage[switch]{lineno}
\usepackage{lipsum}

\makeatletter
\def\ps@IEEEtitlepagestyle{
  \def\@oddfoot{\mycopyrightnotice}
  \def\@evenfoot{}
}
\def\mycopyrightnotice{
  {\footnotesize This work has been submitted to the IEEE for possible publication. Copyright may be transferred without notice, after which this version may no longer be accessible.\hfill} 
  \gdef\mycopyrightnotice{}
}

\begin{document}
\title{
A Structurally Regularized CNN Architecture via Adaptive Subband Decomposition
%A Parallel CNN Architecture with an Adaptive Subband Decomposition Front-End
}
%ZZ: title too long&indirect, how about "Sub-band Decompotions for Structurally Regularized CNN Architecture"

\author{Pavel~Sinha,~\IEEEmembership{Student Member,~IEEE,}
        Ioannis~Psaromiligkos,~\IEEEmembership{Member,~IEEE,}
        and~Zeljko~Zilic,~\IEEEmembership{Senior Member,~IEEE}% <-this % stops a space
\IEEEcompsocitemizethanks{\IEEEcompsocthanksitem P. Sinha, I. Psaromiligkos and Z. Zilic  are with the Department
of Electrical and Computer Engineering, McGill University, Montreal, Canada.\protect\\
E-mail: pavel.sinha@mail.mcgill.ca,  ioannis.psaromiligkos@mcgill.ca and zeljko.zilic@mcgill.ca
}}
%%%%%%%%%% Notations used %%%%%%%%%% 
% To be consistent, notations used:
% (All notations in italics)
%-----------------------------------
% K = Total no. subbands/CNNs
% N = Total no. of FC layers
% I = Total no. of CNN layers
% M = Total no. of DWT layers
%-----------------------------------
% k = Index of subband
% n = Index of FC layer
% i = Index of CNN layer
% m = Index of DWT layers
% l = Index of iteration in equations 7 & 8
%-----------------------------------
% f(i)   = No. of filters in layer 'i' of CNN
% f(i-1) = No. of input channels at the 'i'th CNN layer
% s(i)   = Spatial size of filter
% p(i)   = Spatial size of the output feature map
%-----------------------------------
% BCNN	    Baseline CNN
% SSR-CNN	Single channel subband regularized CNN
% MSR-CNN	Multi channel subband regularized CNN
% WSD	    Wavelet subband decomposition
% CASD	    Constrained adaptive subband decomposition
% ASD	    Adaptive subband decomposition
%-----------------------------------
%\linenumbers

\IEEEtitleabstractindextext{%
\begin{abstract}
We propose a generalized convolutional neural network (CNN) architecture that first decomposes the input signal into subbands by an adaptive filter bank structure, and then uses convolutional layers to extract features from each subband independently. 
Fully connected layers finally combine the extracted features to perform classification.
The proposed architecture restrains each of the subband CNNs from learning using the entire input signal spectrum, resulting in structural regularization.
Our proposed CNN architecture is fully compatible with the end-to-end learning mechanism of typical CNN architectures and learns the subband decomposition from the input dataset.
We show that the proposed CNN architecture has attractive properties, such as robustness to input and weight-and-bias quantization noise, compared to regular full-band CNN architectures.
Importantly, the proposed architecture significantly reduces computational costs, while maintaining state-of-the-art classification accuracy.

Experiments on image classification tasks using the MNIST, CIFAR-10/100, Caltech-101, and ImageNet-2012 datasets show that the proposed architecture allows accuracy surpassing state-of-the-art results.
On the ImageNet-2012 dataset, we achieved top-5 and top-1 validation set accuracy of 86.91\% and 69.73\%, respectively.
Notably, the proposed architecture offers over 90\% reduction in computation cost in the inference path and approximately 75\% reduction in back-propagation (per iteration) with just a single-layer subband decomposition.
With a 2-layer subband decomposition, the computational gains are even more significant with comparable accuracy results to the single-layer decomposition.
\end{abstract}
%ZZ: above abstract doesn't show enough novelty vs. conf. paper

}

\maketitle

\IEEEdisplaynontitleabstractindextext

\IEEEpeerreviewmaketitle

\section{Introduction}\label{sec:introduction}

\IEEEPARstart{D}eep Learning has revolutionized the fields of object recognition~\cite{NIPS2012_4824}, semantic segmentation~\cite{long2015fully}, image captioning~\cite{vinyals2015tell}, human pose estimation~\cite{Toshev_2014} and more.
Convolutional neural networks (CNNs) capable of learning complex hierarchical feature representations for handwriting recognition were first reported in~\cite{Lecun98gradient-basedlearning}.
Since that seminal work, several improvements have been made to the CNN architecture, including 
Alexnet~\cite{NIPS2012_4824}, 
ZFNet~\cite{zeiler2013visualizing},
VGG~\cite{simonyan2015deep},
GoogleNet~\cite{szegedy2014going},
ResNet~\cite{he2015deep},
Spatial Transform Network~\cite{jaderberg2016spatial},
Inception network~\cite{DBLP:conf/cvpr/SzegedyLJSRAEVR15},
Spatial Pyramid Pooling~\cite{DBLP:journals/pami/HeZR015},
Siamese Network~\cite{Koch2015SiameseNN},
SqueezeNet~\cite{iandola2016squeezenet}, 
VGGFace~\cite{Parkhi2015DeepFR}, 
VGGFace2~\cite{cao2018vggface2},
FaceNet~\cite{Schroff_2015},
Transformer in CNN~\cite{liu2022vision},
and more. 

In general, CNN training is vulnerable to overfitting due to a large number of weights.
%This overfitting is not completely eliminated by increasing the size of the dataset~\cite{poggio2017deep}.
%Hence, many articles address overfitting issues.
%Some of the recent techniques try to attenuate the impact of overfitting by including data augmentation~\cite{NIPS2012_4824},  dropouts~\cite{NIPS2012_4824}, and cutouts~\cite{devries2017improved}.
An approach  that has been extensively studied even before deep neural networks is regularization.
It was first introduced in~\cite{NIPS2009_3848}, where adaptive regularization of weight vectors were shown. %also shown. %further shown in~\cite{NIPS2009_3848}.
%An algorithm for regularized training that uses a modified conjugate gradient method to determine the hidden weights has been described in~\cite{7966069}.
Another regularization approach puts constraints on the structure of the network.
A theoretical study on structure-based overfitting was presented in~\cite{sun2015structure}. 
In recent work, a Graph-Spectral-based regularization method was introduced in~\cite{2018arXiv181000424T}.
A wavelet-regularized semi-supervised learning algorithm was proposed in~\cite{6736905} using spline-like graph wavelets.

In the last two decades, significant research has been done on subband analysis, and its advantages are now well-established. 
It is not surprising that wavelets have been used in conjunction with ML methods in a variety of classification tasks. 
As an example from the pre-deep learning era, in \cite{doi:10.1117/12.208730} a wavelet transform is applied to the input signal before being processed by a single-layer neural network.
In~\cite{Kang_2017} and~\cite{7838150}, the input signal is decomposed using a wavelet transform, followed by processing of each subband with a CNN.
In these works, a hierarchical wavelet-based subband decomposition is used that decomposes only the low-frequency components into further subbands, thereby giving higher importance to the low-frequency subbands and combines the high-frequency components into a single subband.
In~\cite{fujieda2018wavelet} a similar wavelet-based decomposition giving importance to the low-frequency subbands is proposed, where the output at each decomposition layer is merged back to the high-frequency path, thus maintaining the full spectrum in the signal flow path.
A similar approach can be found in~\cite{ulicny2022harmonic}, where a discrete cosine transform is performed on the input image followed by a CNN network. 
In~\cite{Oyallon_2018}, the input to the CNN is compressed with a first-order scattering transform, and is demonstrated theoretically that the compression is able to retain features necessary for classification by the CNN.
A wavelet scattering network is presented in~\cite{bruna2012invariant} that computes a translation invariant image representation that is stable to deformations and preserves high-frequency information for classification. 

In our previous work~\cite{8804202}, we proposed the Wavelet Subband Decomposition (WSD) structure 
that decomposes an input image using wavelets and then processes each of the resulting subbands using separate convolutional layers whose outputs are combined by a fully connected (FC) layer. 
As we discuss in~\cite{8804202}, the network exhibits several attractive properties such as structural regularization, immunity from input and weight quantization noise, etc.

A common characteristic of the above methods is that the subband decomposition front end uses a fixed wavelet kernel, while the learning process only involves the CNN part following the subband decomposition. 
Therefore, it is likely that the decomposition structure is sub-optimal.  
This begs the question: can the end-to-end training process include the subband decomposition front-end, and if so, what are the benefits in performance or complexity? 

Motivated by this question, we investigate learning the subband decomposition of the input signal from the dataset.
Specifically, we study two subband decomposition structures: Constrained Adaptive Subband Decomposition (CASD) and Adaptive Subband Decomposition (ASD). 
In CASD, the subband decomposition filter weights are constrained to result in non-overlapping subbands. 
On the other hand, in ASD, no such constraint is imposed on the filter weights. 
The training of the weights in either structure is done by back-propagating the error derivatives from the CNN to the subband decomposition filter structures, making them fully trainable end-to-end. 
%We explore this by %first 
%having the subband decomposition structure learned from the dataset and maintaining a complementary overlapping subband decomposition structure, as presented in the previous work. Following this, the subband decomposition structure could learn from the dataset and have unconstrained overlapping subband decomposition that best optimizes the cost function.
%Motivated by this question, in this paper, we build upon our previous work~\cite{8804202}, and we propose to decompose the input into subbands using linear filters whose weights are learned in an end-to-end training process that integrates the learning of the subband decomposition filter weights with the process of CNN training.
%This is done by propagating the error derivatives from the CNN to the subband decomposition filter structures. 
In later sections, we study and compare the characteristics of both ASD and CASD.  
%The constraint mechanism in the CASD structure enables non-overlapping subbands which are computationally efficient compared to ASD.

Further, we consider three CNN architectures. The first is the Multi-Channel Subband Regularized CNN (MSR-CNN) which processes each of the decomposed subband channels with a separate CNN. 
The second architecture, Single-Channel Subband Regularized CNN (SSR-CNN), stacks up the subbands into multiple  channels forming a single input to a single CNN. 
Each of the SSR-CNN and MSR-CNN architectures includes an FC layer that performs classification by combining the features extracted from the CNN(s). Finally, we consider a regular full-band CNN, Baseline CNN (BCNN), that computes the subband decomposition of the input and processes it through a single regular CNN. We note that SSR-CNN and BCNN serve mainly as benchmark architectures against which we evaluate the performance of MSR-CNN.

To demonstrate the performance of the considered architectures, we apply them in an image classification problem. 
%This is done solely for illustration purposes. 
The proposed structures can be applied to other kinds of problems, such as audio and communication signal processing.
We show that our proposed architecture with the ASD front-end structure and MSR-CNN architecture achieves accuracy comparable to, if not better, than the state of the art, and it generalizes much better than a standard CNN model. 
Importantly, we achieve these results by requiring less than 10\% of the computations in the inference path and only 25\% of the computations in backpropagation compared to a full-band standard CNN.
%In our experimentation, the subband decomposition with ASD followed by MSR-CNN is found to perform the best.
An additional benefit is that our proposed subband decomposition structure is a form of structural regularization. 
This can be attributed to the fact that, in the case of MSR-CNN, the presented method inhibits any CNN from training on information available in other subbands. 
This provides additional flexibility to reach a trade-off between restricting the scope of each CNN for optimal structural regularization vs. enabling each of the CNNs to extend its scope across the subbands which may benefit feature extraction.
Further, each CNN is subject to weight regularization. 
Combined, the structural and weight regularization, lead to reduced over-fitting.
Overall, as our experimental studies show, the proposed subband architectures offer significant computational reduction with minimal impact on performance.

The paper is structured as follows: In Section~\ref{sec:proposed_architecture}, we describe the proposed subband decomposition structures and associated CNN architectures. 
In Section~\ref{sec:Properties_of_the_Proposed_Architecture}, we study the properties of the proposed architectures, followed by a description of the experimental setup and results in Section~\ref{sec:Experimental_Setup_and_Results}. 
Finally, Section~\ref{sec:Conclusion} concludes the paper.

\section{Proposed Architecture}
\label{sec:proposed_architecture}

\begin{figure}[t]
    \centering
    \includegraphics[scale=0.68]{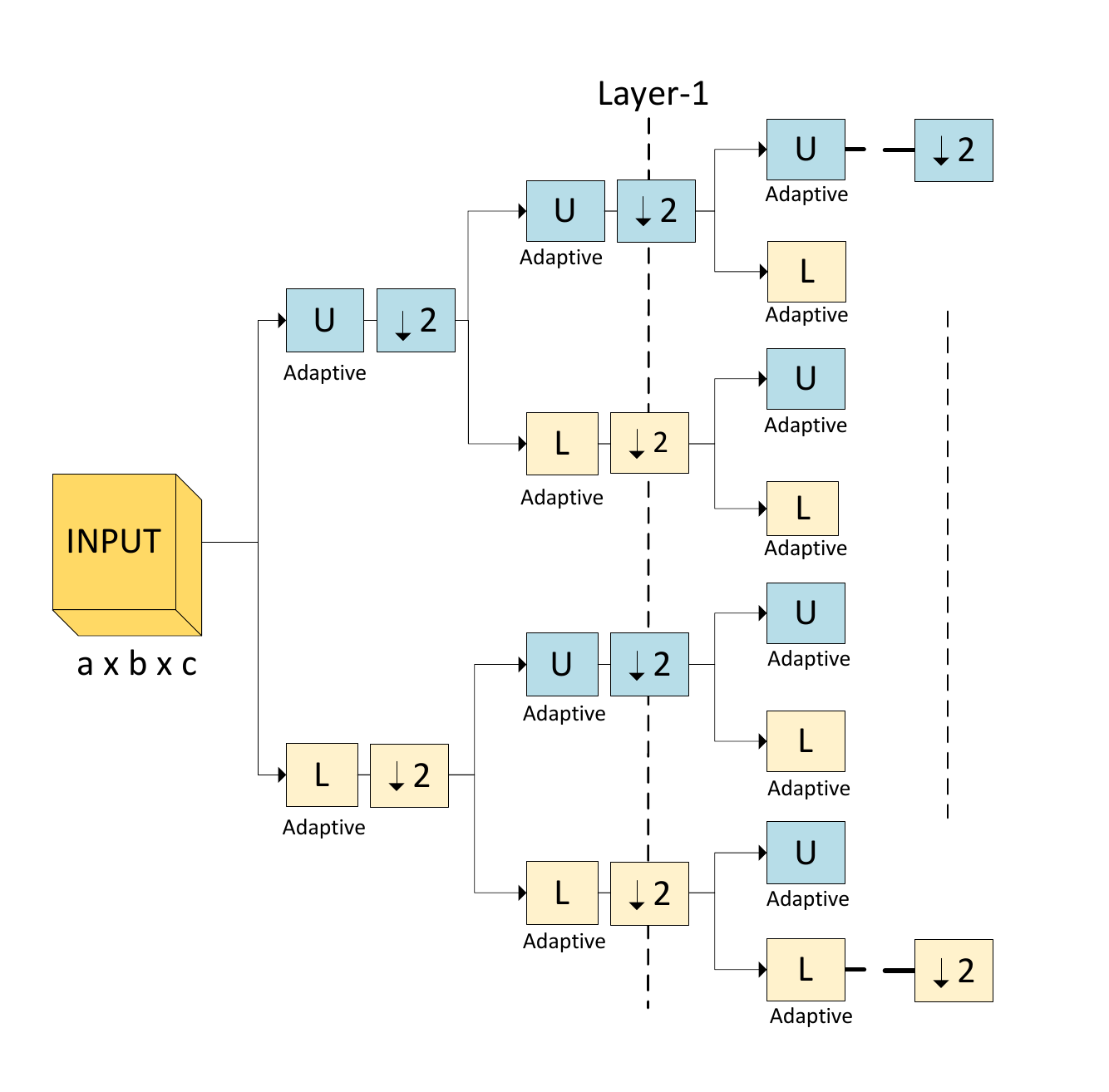}
    \caption{ASD filter structure with $M$-Layer decomposition, where $U$ and $L$ correspond to the upper and lower part of the subband decomposition linear filter structure. Please note that  $U$ and $L$ do not necessarily correspond to low pass and high pass filters, respectively, but only  distinguish the two branches of each subband decomposition.}
    \label{adaptive_filter}
\end{figure}

%In Fig.~\ref{adaptive_filter}, we show the proposed ASD structure. 
%Similar to $HP$ and $LP$ in the case of WSD in~\cite{8804202}, the $U$ and $L$ blocks represent finite impulse response (FIR) linear decomposition filters, each followed by a decimation-by-2 block. 
%In contrast to WSD, however, where the blocks $HP$ and $LP$ correspond to static complementary high-pass and low-pass filters, respectively, the filters $U$ and $L$ are learned via end-to-end training and may not be high-pass or low-pass filters nor spectrally complementary. 
 
\begin{figure}[t]
    \centering
    \includegraphics[scale=0.65]{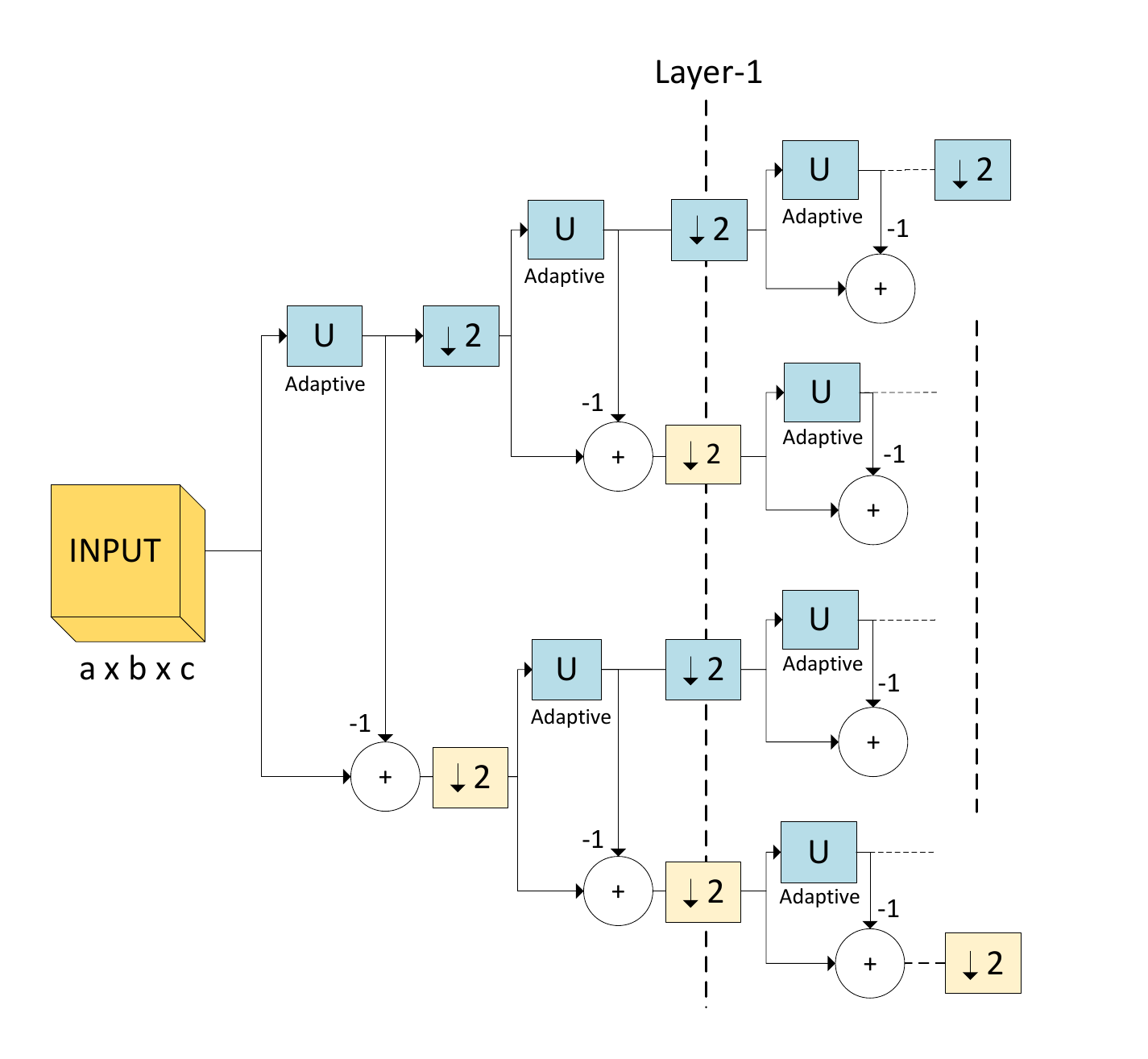}
    \caption{CASD filter structure with $M$-Layer spectral complementary subband decomposition. The structure resembles the ASD structure with the constraint that the lower branch filter of each sub-band decomposition is $1-U$, where $U$ is the upper branch filter. 
    }
    \label{loss_less_adap_tran}
\end{figure}

In this section, we describe the proposed frontends (ASD and CASD) and the MSR-CNN and SSR-CNN architectures in more detail. 

\subsection{Proposed Subband Decomposition Structures}
\label{subsection:Proposed_Subband_Decomposition_Structures}
\subsubsection{Adaptive Subband Decomposition (ASD)}
\label{sec:ASD}
Fig.~\ref{adaptive_filter} depicts the  ASD structure with $M$ layers. 
Similar to $HP$ and $LP$ in  WSD~\cite{8804202}, the $U$ and $L$ blocks represent decomposition filters.
Specifically, each block comprises a collection of $c$ filters operating in parallel on the $c$ channels of the input. The filter for the $i$th channel, $i=1,\ldots, c$, is a real 2-dimensional filter of order $(2N+1)\times (2N+1)$, where $N$ is an integer, with impulse response denoted by $W(k,l;i)$. 
Representing  the input by a tensor of dimensions $a\times b \times c$, $X(m,n;i)$, the output tensor $Y(m,n;i)$ is given by the convolution sum
\begin{align}
Y(m,n;i) &= \sum_{k=-N}^{N}{\sum_{l=-N}^{N}{W(k,l;i)X(m-k,n-l;i)}}
\label{eq_fir}
\end{align}

The $U$ and $L$ blocks are followed by decimation-by-2 along the $x$ and $y$ dimensions only.
Hence, at each layer, each channel of the input is split into four subbands with half the resolution along each of the $x$ and $y$ dimensions.
%The input is represented by a tensor of dimensions $a\times b \times c$, where $c$ is the number of input channels, first decomposed into $4$ subbands, by a filtering-and-decimation process along the rows and columns. 
%Each subband is represented by a tensor of dimensions $a/2\times b/2 \times c$, where $a$, $b$, and $c$ are the number of rows, columns, and channels of the original input.
This filtering-and-decimation process gets repeated for each subband as we move to the next decomposition layer. 
For example, for an input of size 224$\times$224 with 3 channels, i.e., 224$\times$224$\times$3, an a 1-layer subband decomposition provides four subbands with dimensions 112$\times$112$\times$3. 
For 2-layer decomposition, the same input yields 16 subbands of dimensions 56$\times$56$\times$3 each.
We note that the net dimension of the input and the final stack of the generated subbands are the same.

%In contrast to WSD, where the blocks $HP$ and $LP$ correspond to static complementary high-pass and low-pass filters, respectively, the filters $U$ and $L$ are learned via end-to-end training and may not be high-pass or low-pass filters nor spectrally complementary. 
The network is end-to-end trained by back-propagating the error gradient up to the subband structures.
It is, therefore, possible that, in contrast to WSD, the subbands may not be orthogonal leading to an overall lossless decomposition of the input signal. 
It is interesting to note that, as we show later, the subband structure learns filters of known types, such as band-pass, band-stop, low-pass, and high-pass structures, typically used in signal processing. 

%Fig.~\ref{adaptive_filter} shows a generalized $M$-layer ASD structure. 

\begin{figure}[t]
    \centering
    \includegraphics[scale=0.75]{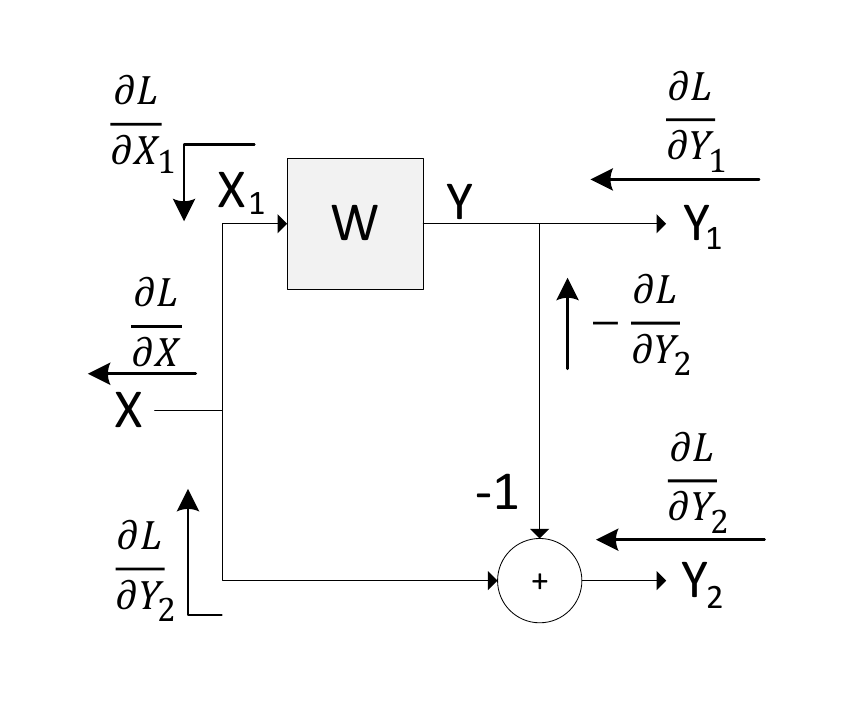}
    \caption{Flow of error derivative during back-propagation in the CASD subband decomposition structure. 
    %We show the backward flow of the error derivatives along the data flow path. The convolution operation in the figure is represented by Eq.~\ref{eq_fir}.
    %The weights $W$ are updated at training based on error derivative $\frac{\partial L}{\partial W}$. 
    The backpropagated error derivatives coming from the next decomposition stage are given by $\frac{\partial L}{\partial Y_{1}}$ and $\frac{\partial L}{\partial Y_{2}}$, while the error derivative passed on to the previous stage is $\frac{\partial L}{\partial X}$.
    }
    \label{optimal_filter_2}
\end{figure}

\subsubsection{Constrained Adaptive Subband Decomposition (CASD)}
We can obtain a lossless subband decomposition that preserves the net information content of the input signal by further constraining the $U$ and $L$ filters into being spectrally complementary. 
The resulting CASD structure is shown in Fig.~\ref{loss_less_adap_tran}. 
As we can see, CASD is a special case of ASD with $L=1-U$, i.e., $U$ has the same structure as in the case of ASD described in Section \ref{sec:ASD}. 
%Fig.~\ref{loss_less_adap_tran} depicts a generalized $M$-Layer spectrally complementary subband decomposition structure, where block $U$ corresponds to the subband decomposition filter implemented by a $(2N+1)\times (2N+1)$ 2-dimensional filter with real coefficients. 
The input-output relationship is given by (\ref{eq_fir}) with $W=U$.
Similar to ASD, the network is end-to-end trained. 

One of the benefits of the CASD structure is that it reduces the number of filter weights and, therefore, the computational cost of the subband decomposition structure at training by half; the cumulative effect of this reduction across training iterations can be significant. Another benefit of CASD is a higher degree of regularization compared to ASD which, depending on the dataset, can be beneficial. 
As we will see in the experimental results section, CASD (worse performance but also lower computational complexity) allows us to achieve a different tradeoff point than ASD which could be preferable depending on the application.

\subsubsection{Training of front-end structures}
Note that the decomposition structures shown in Fig.~\ref{adaptive_filter} and Fig.~\ref{loss_less_adap_tran} are regular convolution filter structures, and hence the standard back-propagation mechanism is applicable. We illustrate this by deriving the back-propagation equation for the CASD structure. 
Referring to Fig.~\ref{optimal_filter_2}, let $L$ be the loss function of the overall CNN, $X$ be the input, and $Y_1$ and $Y_2$ be the two outputs going to the next stage of the decomposition structure. Here $Y_1$ and $Y_2$ are complementary to each other in terms of spectral content~\cite{Wanhammar1999DSPIC},~\cite{5339431}, ([c. 4, p. 125)].  
Also, $\frac{\partial L}{\partial Y_i}$ is the back-propagating error derivative wrt the output $Y_i$, $i\in\{1,2\}$, $\frac{\partial L}{\partial W}$ the error derivative wrt the weights $W$, $\frac{\partial L}{\partial X}$ the error derivative wrt the input $X$ or the gradient passing down to the downstream module and $\frac{\partial Y}{\partial X_1}$ the local gradient of the filter $W$. 
Then, we can write
\begin{align}
\frac{\partial L}{\partial Y} &= \frac{\partial L}{\partial Y_1} - \frac{\partial L}{\partial Y_2}
\label{eq_1a} \\
\frac{\partial L}{\partial X} &= \frac{\partial L}{\partial X_1} + \frac{\partial L}{\partial Y_2}~= \frac{\partial L}{\partial Y}\frac{\partial Y}{\partial X_1} + \frac{\partial L}{\partial Y_2} \nonumber\\
&= \frac{\partial Y}{\partial X_1}(\frac{\partial L}{\partial Y_1} - \frac{\partial L}{\partial Y_2}) + \frac{\partial L}{\partial Y_2}
\label{eq_2a}\\
\frac{\partial L}{\partial W} &= \frac{\partial L}{\partial Y}\frac{\partial Y}{\partial W}~= \frac{\partial Y}{\partial W}(\frac{\partial L}{\partial Y_1} - \frac{\partial L}{\partial Y_2})
\label{eq_3a}
\end{align}
Computing $\frac{\partial L}{\partial X}$ is needed to propagate the error derivative while $\frac{\partial L}{\partial W}$ is required to update the filter weights.

\subsection{CNN Architectures}
\begin{figure}[t]
    \centering
    \includegraphics[scale=0.57]{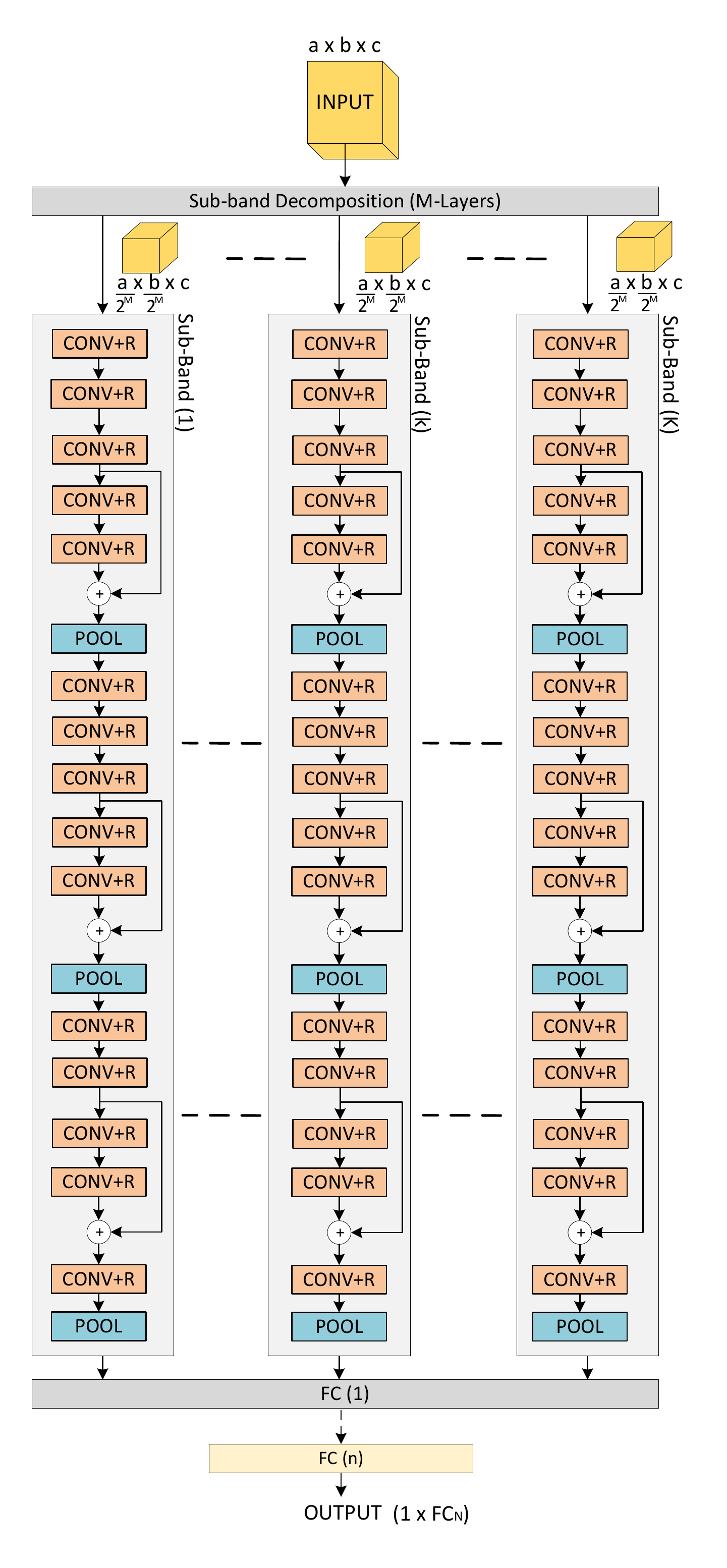}
    \caption{Architecture of the $M$-Layer MSR-CNN, parameterized by input dimensions (a$\times$b$\times$c), $K=2^M$ number of subbands, $I$ convolutional layers per subband, $N$ FC layers and ${\text{FC}}_{N}$ output classes, all open to optimization. The subband decomposition block can be either of the earlier structures: ASD, CASD or WSD. The CONV+R modules represent a convolution operation followed by a ReLu.}
    \label{msr-cnn}
\end{figure}

\begin{figure}[t]
    \centering
    \includegraphics[scale=0.62]{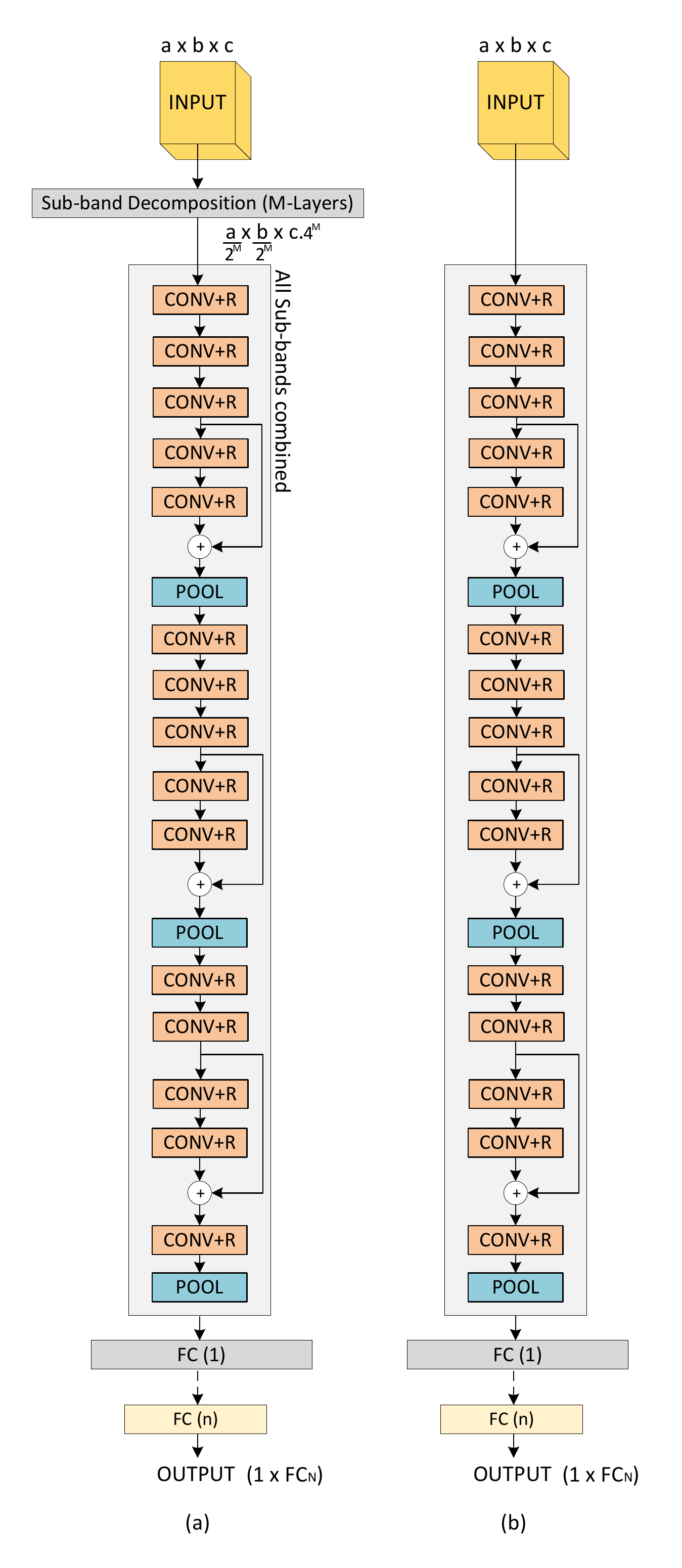}
    \caption{(a) Architecture of the $M$-Layer wavelet decomposed SSR-CNN - any of the ASD, CASD or WSD subband decomposition methods can be used. (b) The architecture of BCNN.}
    \label{ssr-cnn}
\end{figure}
\label{subsection:CNN_Architectures_with_Subband Decompositions}
As described earlier, the subbands are processed by individual CNNs to extract relevant features before passing them through the fully connected (FC) layer for classification. The overall architecture called  the MSR-CNN, is shown in Fig.~\ref{msr-cnn}.  
%The architecture restricts the field of view of each sub-band CNN to a dedicated subband, making each CNN independent from the rest of the subbands. 
As a benchmark, we also consider a second architecture that passes the complete subband decomposition through a single CNN, as shown in Fig.~\ref{ssr-cnn}(a). 
We call this architecture the SSR-CNN.
Fig.~\ref{ssr-cnn}(b) shows a second benchmark architecture, the Baseline CNN (BCNN). The BCNN architecture is similar to the MSR-CNN and the SSR-CNN in terms of the number of convolutional stages and the total FC layers, except the BCNN does not have the subband decomposition structure. Instead, the input image is passed directly into the CNN. 

%From the subband decomposition structure, it becomes clear that the input dimension per subband gets substantially reduced and the output from the subband decomposition structure becomes a collection of low-resolution versions of the input representing various subbands. 
%However, one should not confuse this with the functioning of a pooling layer in a CNN or conclude that this could potentially replace a pooling layer or other parts of the CNN. 
%Though the pooling layer effectively reduces the dimension of the input by half on both its length and breadth, due to the hierarchy of convolutional layers prior to a pooling layer, important features from the input peculate through each layer. 
%After subsequent layers, we see a concentration of high-level features that help in partitioning the hyperspace by the FC layers to form the classification results. 
%These high-level features are learned as a result of back-propagating the error, which reduces the cost function during training.

\subsubsection{Multi-Channel Subband Regularized CNN (MSR-CNN)}
\label{sec:ssrcnn}
Fig.~\ref{msr-cnn} shows the generalized $L$-level MSR-CNN architecture in which the subbands are processed separately by individual CNNs, each with $I$ layers. This restricts the field of view of each CNN, making them indifferent to the rest of the subbands.
%The decomposed subbands are obtained using any of the three subband decomposition structures (ASD, CASD, WSD), which includes decimation along the rows and columns.
%In the case of WSD, the filter coefficients are fixed.
%In ASD and CASD, the filter coefficients are learned through back-propagation, as explained earlier, with the error derivatives propagated from the subband convolutional layers all the way to the input.
The convolutional layers are similar to that of Alexnet~\cite{NIPS2012_4824} and/or VGG16~\cite{simonyan2015deep}, with the difference of having a residual layer connection~\cite{he2015deep} before the pooling layer, as per Fig.~\ref{msr-cnn}. 
The addition of the residual layers, inspired by ResNet~\cite{he2015deep}, significantly helps in speeding up training by allowing the error gradients to easily propagate to the initial layers.
Finally, $N$ FC layers combine the feature outputs of the subband CNNs and perform classification. 

We note that even though the input to the subband CNNs is already reduced in dimension, we still need pooling layers in the CNN to extract the dominant features per subband.
This is done with a hierarchy of convolutional layers followed by a pooling layer that keeps the outputs with the largest magnitude, discarding the rest.
In contrast, the decimation-by-2 blocks in the subband decomposition structure drop every 2nd filter output sample along the x and y dimensions, regardless of their magnitude.
Therefore, the primary function of the wavelet subband decomposition is not to do dominant feature extraction. 
Rather, it assists in the extraction of the dominant features by the CNNs that follow.

\subsubsection{Single-Channel Subband Regularized CNN (SSR-CNN)}
Fig.~\ref{ssr-cnn}(a) shows the general architecture of SSR-CNN.
The SSR-CNN architecture shares the same front-end subband decomposition structure as the MSR-CNN, i.e., ASD, CASD, or WSD.
An $M$-layer subband decomposition structure, with input tensor of dimensions $a\times b \times c$, where $a$, $b$, and $c$ are the number of rows, columns, and channels, will result in ${4}^{M}$ total subbands, each  represented by a tensor of dimensions $(a/2^M)\times (b/2^M) \times c$.
In contrast to the MSR-CNN architecture, all the channels of the subbands are stacked together and treated as a single input of dimension $((a/2^M)\times (b/2^M) \times (c\cdot 4^M)$) to a single CNN, followed by a fully connected layer to classify.

\begin{figure*}%[hbt]
\includegraphics[trim=5.0cm 0 0 0, scale=0.45]{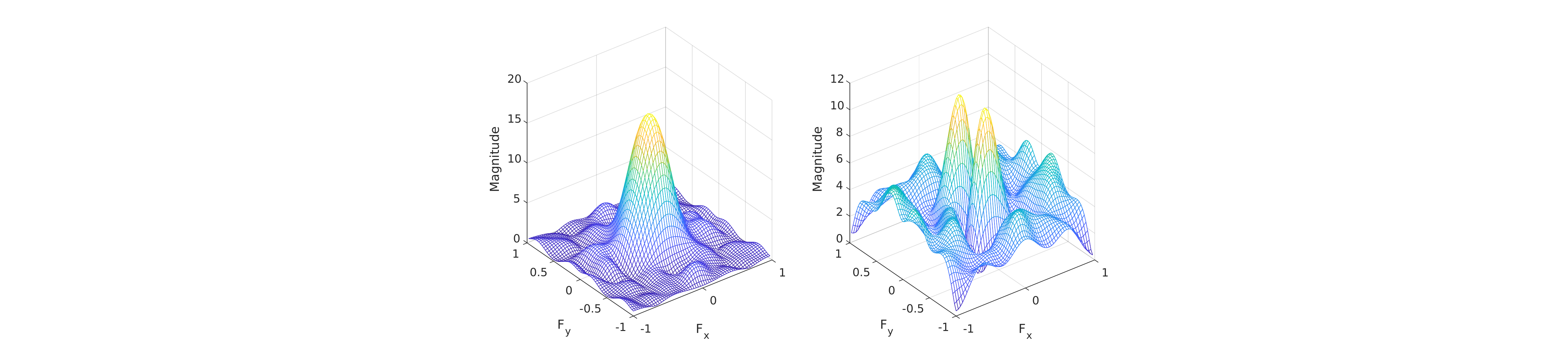}
\centering
\end{figure*}
\begin{figure*}%[hbt]
\includegraphics[trim=5.0cm 0 0 0, scale=0.45]{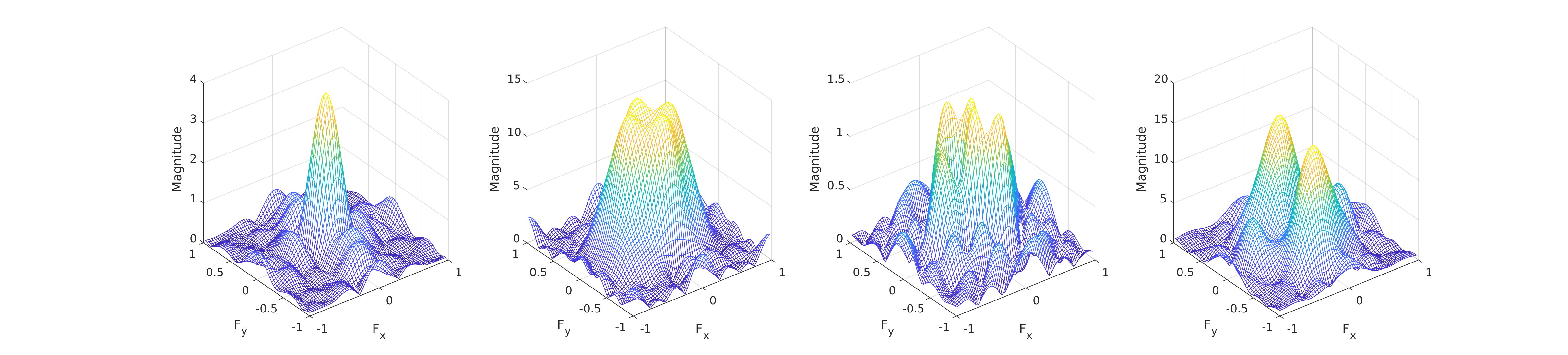}
\centering
\caption{
A 2D frequency response of layer-1 decomposition of the ASD filter structure.
The spatial frequencies $F_x$ and $F_y$ are normalized to $\pi$ rads/sample. 
The top row corresponds to the filter frequency response that decomposes the input into the first upper and lower paths in Fig~\ref{adaptive_filter}.
The second row corresponds to the subsequent upper and lower path decomposition filter frequency responses after the first upper and lower path decomposition filters. 
This completes the decomposition of layer-1 of the input into four subbands.
It can be seen that the filter frequency response of the decomposition structure is formed out of known filter responses in the literature such as low-pass, band-pass, band-reject, and high-pass frequency responses.
}
\label{fig_filt_resp}
\end{figure*}

\begin{figure}[t]%{\textwidth}
\vstretch{0.82}{\includegraphics[scale=0.25]{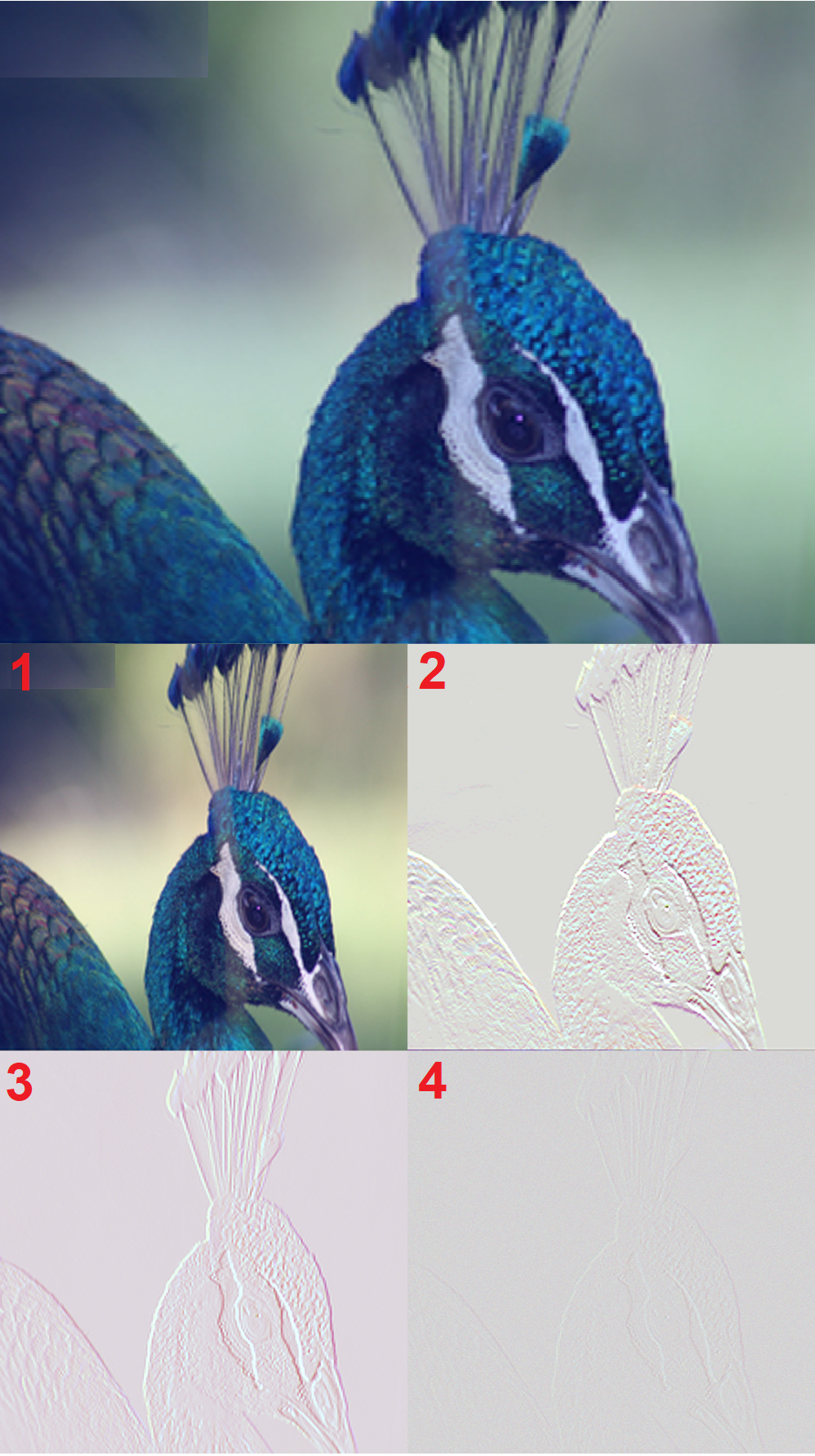}}
\centering
\caption{
Top row: Input image. Second and third rows: Images numbered 1 through 4, correspond to the LL, LU, UL, and UU subbands, respectively, obtained using the Daubechies (D2) family of basis functions~\cite{57199}. 
%As we can see, most of the low-frequency contents are present in the low-low frequency subband. 
%The other subbands mostly contain high-frequency contours of the input image.
}
\label{fig_in_img_DWT}
\end{figure}
\begin{figure}[bt!]%{\textwidth}
\vstretch{0.80}{\includegraphics[scale=0.125]{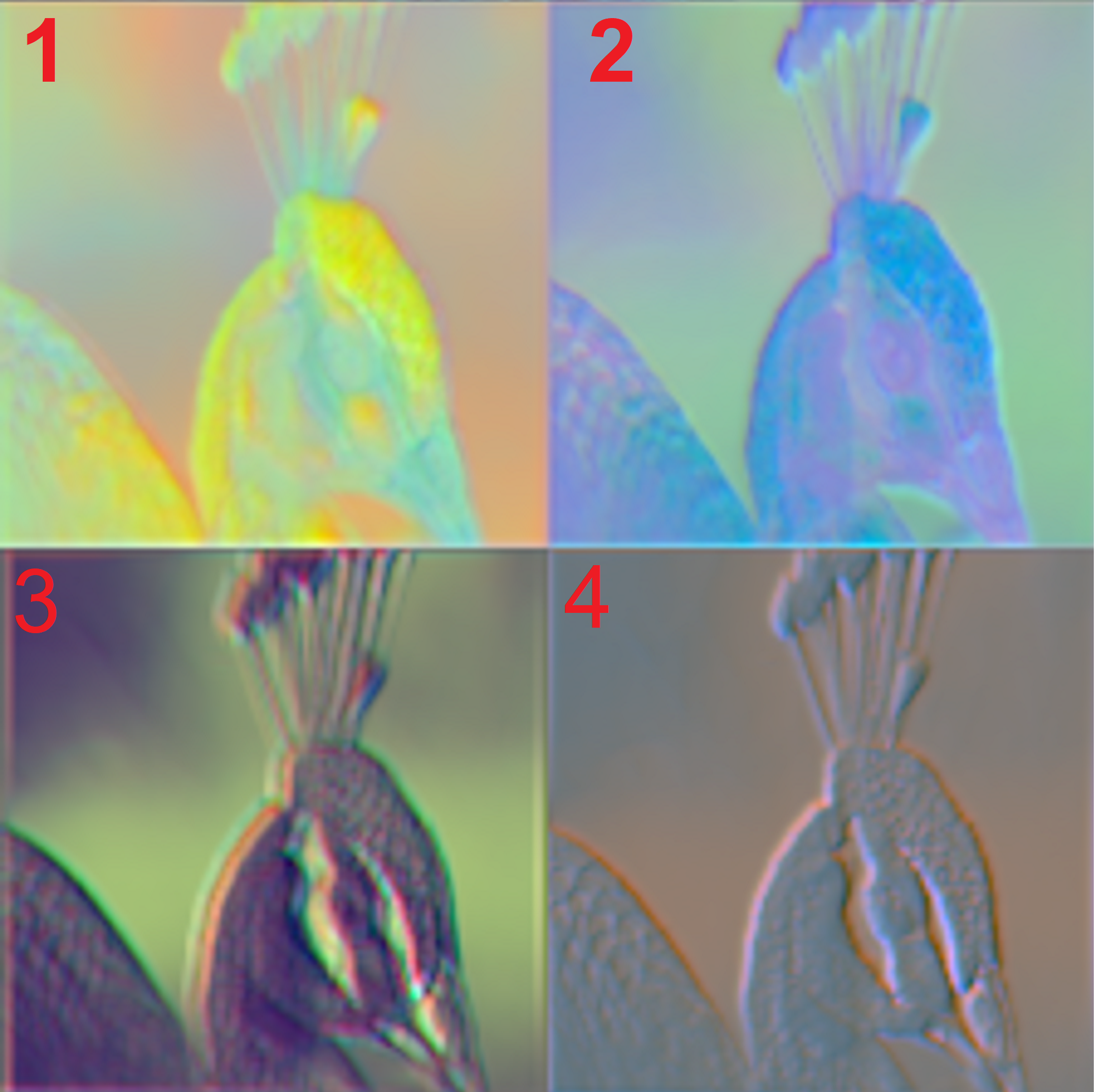}}
\centering
\caption{
%Input image (top row) decomposed into four subbands using ASD in Fig.~\ref{adaptive_filter}. 
Images numbered 1 through 4, correspond to the LL, LU, UL, and UU subbands, respectively, for the same input image as in Fig.~\ref{fig_in_img_DWT}, obtained using the ASD decomposition structured trained on the Imagenet dataset.
%As can be seen, all four channels contain a mix of low and high-frequency components unlike that of the subbands decomposed using DWT in Figure~\ref{fig_in_img_DWT}.
}
\label{fig_in_img}
\end{figure}

\begin{figure*}[bt!]%{\textwidth}
\includegraphics[scale=0.34]{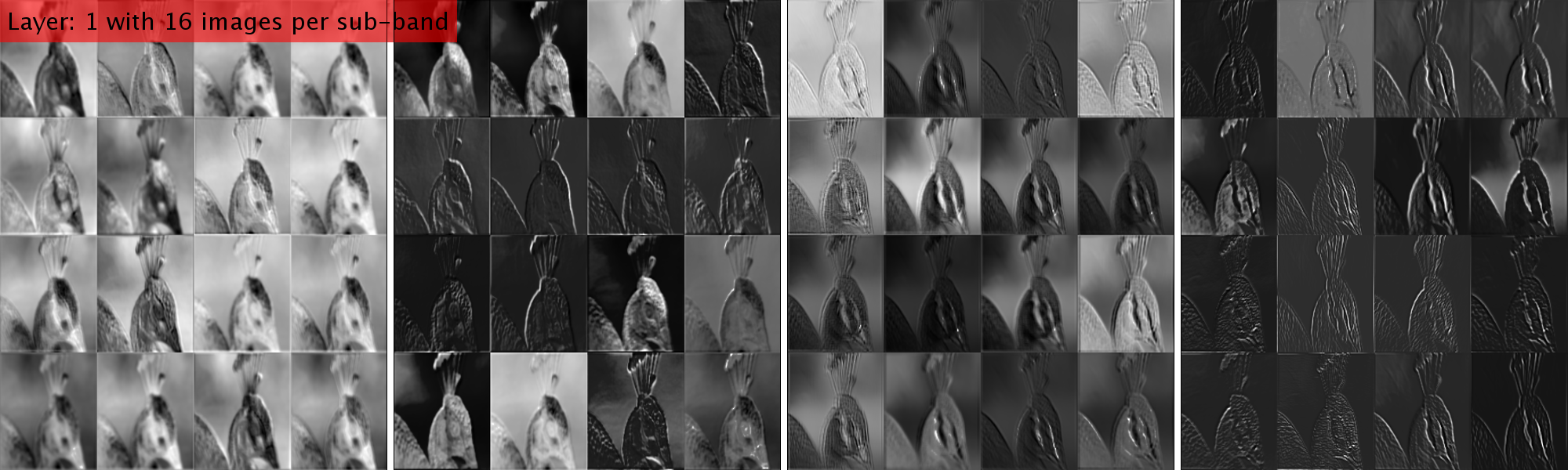}
\centering
\caption{
The activation map of the first convolutional layer output of MSR-CNN with a 1-layer ASD structure that decomposes the input into 4 subbands.
Each of the subband CNNs in the first convolution layer has 16 filters, i.e., the layer outputs 16 channels per subband. 
Each block of 16 images shown corresponds to one of the four subbands LL, LU, UL, and UU (shown in Fig.~\ref{fig_in_img}) from left to right. 
}
\label{output_layer_1}
\end{figure*}

\begin{figure*}[bt!]%{\textwidth}
\includegraphics[scale=0.34]{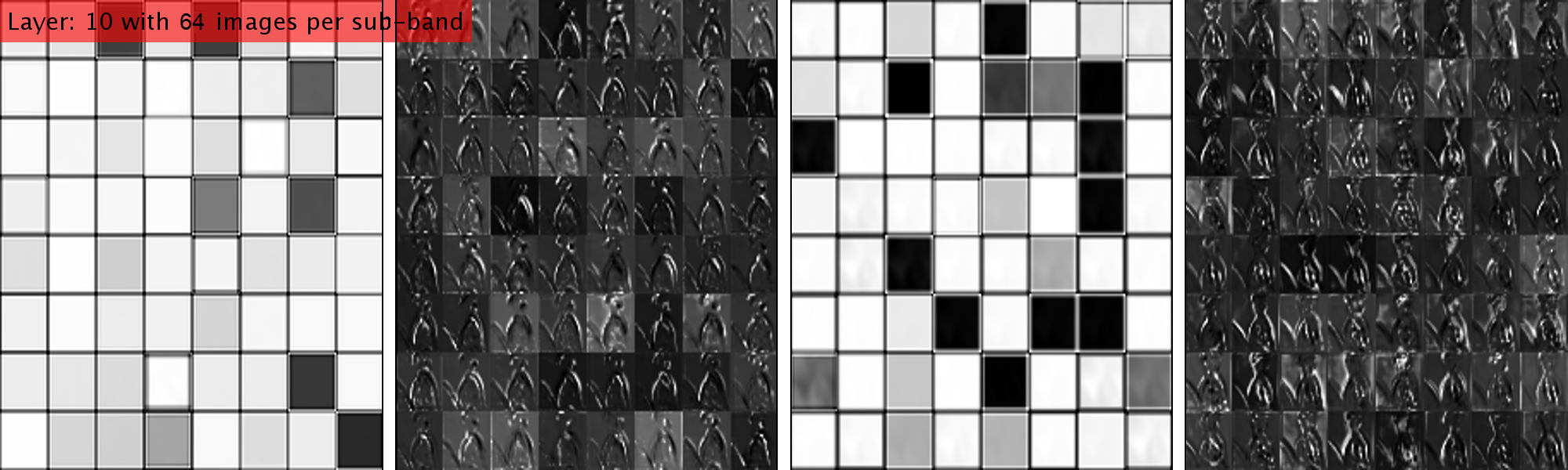}
\centering
\caption{
Activation map of the 10th convolutional layer of MSR-CNN with a 1-layer ASD structure that decomposes the input into 4 subbands.
Each of the subband CNNs at the 10th convolution layer has 64 filters, i.e., the layer outputs 64 channels per subband. 
Each block of 64 images corresponds to one of the four subbands LL, LU, UL, and UU (shown in Fig.~\ref{fig_in_img}) from left to right, respectively.
}
\label{output_layer_10}
\end{figure*}

Both network architectures are end-to-end trained, meaning that the front-end structures (ASD or CASD), CNN and fully connected layers,  are trained together by backpropagating the error derivative from the output layer all the way back to the front-end. 
As we will see, the ASD structure gives better accuracy over both CASD and WSD of~\cite{8804202}, at the expense of higher computational cost.
Furthermore, the CNN architecture with a CASD frontend performs better than with a WSD frontend, as shown in~\cite{8804202}.
The WSD structure offers the least computation cost, while the CASD structure lies between the ASD and the WSD structures, balancing computation cost and performance.

\section{Properties of Proposed Architecture}
\label{sec:Properties_of_the_Proposed_Architecture}
In this section, we discuss several attractive properties of the proposed architecture.
Specifically, we discuss the computational gains and the robustness towards quantization noise this architecture brings.

\subsection{Computational Cost}

%PS: Added this paragraph

The proposed architecture offers a higher degree of parallelism as it allows for the independent processing of each subband.
As we can see from Fig.~\ref{msr-cnn}, we have $K$ CNNs that operate in parallel compared to SSR-CNN or a single CNN in a traditional architecture (cf. Fig.~\ref{ssr-cnn}(b)). 
%This advantage remains true for the backpropagation path as well.
Processing the individual subbands leads to a substantial reduction of the computational cost. 
The computational cost of all convolutional layers in a CNN with $I$ convolutions is $O(\sum_{i=1}^{I} f_{i-1}s_{i}^{2}f_{i}p_{i}^{2})$ \cite{he2014convolutional}, where ${f_i}$ is the number of filters at the $i$th convolution (${f_{i-1}}$ is also referred to as the %number of input channels
input depth
of the $i$th convolution), $s_i$ is the %spatial 
size of the convolution filter and $p_i$ the spatial size of the output %feature map, 
of the $i$th convolution.\footnote{It is assumed that the length and height of the convolution input are equal. If not, $p_i^{2}$  would be replaced by the product of the length and height of the input to the $i$th convolution.}
Subband decomposition reduces the parameter ${p_{i}}$ by half along both length and height for every decomposition layer, significantly reducing the computational cost.
%Subband decomposition of the input signal reduces the input dimensions along rows and columns, each by $2^M$, where $M$ is the number of decomposition layers. 
The total reduction of input dimension along each subband is exponential and is %effectively 
given by $4^M$ for two-dimensional input data such as images. 
%The cost of the convolution operation in a CNN, which accounts for the bulk of computations, depends super-linearly on the size of the convolution filters~\cite{he2014convolutional} and the sample point count per dimension, all of which reduce significantly in our case. 
The reduction in computation complexity also applies to the back-propagation path. 
As we will see in the next section, the reduction in the total computation required for the forward pass and back-propagation over single-iteration is over 98\% and 89\%, respectively. 

Sparsity in both weights and feature maps has motivated the development of different methods to take advantage of the available sparsity and optimize CNN computation requirements~\cite{hoefler2021sparsity},\cite{8771480},~\cite{DBLP:journals/corr/ChangpinyoSZ17}~and~\cite{9034111}.
Typically, the information content of images is distributed unequally over the entire spectrum. 
Therefore, individual subbands generally exhibit higher sparsity compared to the entire spectrum. 
In the proposed subband decomposition structure, this sparsity is introduced at the very input of the subband CNNs. 
This makes it possible to use standard techniques that take advantage of sparsity to reduce CNN complexity, such as~\cite{changpinyo2017power}. 
Finally, the proposed architectures can be combined with one of the numerous techniques to reduce the computational cost and memory requirements of CNNs. For example, by removing redundant kernels as in~\cite{DBLP:journals/corr/LiuWLC17}, or 
 pruning of  kernel coefficients~\cite{DBLP:journals/corr/LiKDSG16}.
%Another  work taking advantage of CNN sparsity to reduce the number of kernels per channel and the number of CNN stages was introduced in.
%SqueezeNet~\cite{DBLP:journals/corr/IandolaMAHDK16} claimed 50$\times$ fewer parameters than AlexNet by using $1\times 1$ convolutional filters.
The reader is referred to~\cite{DBLP:journals/corr/SzeCYE17} for a detailed tutorial and survey of methods to reduce the CNN computational cost.
We note, however, that exploring ways to further reduce the computational cost by applying on of the above techniques is beyond the scope of this work. 

\subsection{Robustness to Input Noise and Quantization Errors}
In practice, the input is subject to quantization noise and corruption due to non-linearities in the capturing device, e.g., lens aberration, incorrect exposure, low lighting condition, and a burst of transmission noise.
Further, in most applications, signals are compressed before storage or transmission.
For instance, image compression takes advantage of the unequal spread of information across the spectrum, allocating fewer bits to parts of the spectrum with less information~\cite{1218191},\cite{1362510}. 
These sources of error ultimately result in noise that is spread unevenly across the spectrum.
The subband decomposition of the input signal enables the network to analyze isolated subbands separately. 
This allows for noise within each subband to be confined within the subband. 
In contrast, in the case of a single full-spectrum input, any source of noise affects the entire spectrum. 

A similar situation arises due to weight and bias quantization, which is imperative for practical implementation, where storage and computation in 64-bit floating-point representation become impossible.
Several papers have studied weight quantization in CNNs, including ~\cite{Lin:2016:FPQ:3045390.3045690}, ~\cite{zhou2017incremental}, and ~\cite{zhou2017adaptive}. 
In CNNs that process the full spectrum of the signal, noise in each coefficient eventually affects the entire spectrum. 
In the proposed CNN architecture, however, weight and noise quantization noise is confined to individual subbands.

%Indeed, as we will see in the next section, all three subband decomposition structures, WSD, CASD, and ASD are highly robust towards input quantization noise compared to the full-band BCNN architecture.

Our experimental studies show that the MSR-CNN is robust to input noise as well as weight and bias quantization noise.
The ASD decomposition structure is shown to be the most robust compared to other subband decomposition structures.
The CASD structure is the second most robust followed by the WSD structure in the third position.
The superior performance of the ASD structure is owed to the ability of the ASD structure to learn from the dataset itself and the flexibility to adapt the subbands to reduce the loss during training without any constraints. 
The performance CASD is slightly lower than that of the ASD structure as a result of the complimentary subband decomposition structure and the restriction that comes along with it.
The performance of the WSD structure indicates that within the constraints of our experimentation, fixed decomposition structures that cannot adapt to the dataset are sub-optimal in nature.

% http://www.tablesgenerator.com/#
\begin{table}[t]
\centering
\caption{Datasets used.}
\label{data_set}
\scalebox{0.83}{
\begin{tabular}{|c|c|c|c|c|}
\hline
\textbf{Dataset} & \textbf{\begin{tabular}[c]{@{}c@{}}No. of training \\ images\end{tabular}} & \textbf{\begin{tabular}[c]{@{}c@{}}No. of test \\ images\end{tabular}} & \textbf{Resolution} & \textbf{Classes} \\ \hline
MNIST~\cite{lecun-mnisthandwrittendigit-2010}            & 60000                                                                      & 10000                                                                  & 28$\times$28               & 10               \\ \hline
CIFAR-10~\cite{Krizhevsky09learningmultiple}         & 50000                                                                      & 10000                                                                  & 32$\times$32               & 10               \\ \hline
CIFAR-100~\cite{Krizhevsky09learningmultiple}        & 50000                                                                      & 10000                                                                  & 32$\times$32               & 100              \\ \hline
Caltech-101~\cite{1384978}      & 6403                                                                       & 2741                                                                   & 300$\times$200             & 102              \\ \hline
ImageNet-2012~\cite{ILSVRC15}    & 1280000                                                                     & 50000                                                                  & Variable            & 1000             \\ \hline
\end{tabular}
}
\end{table}

% http://www.tablesgenerator.com/#
\begin{table*}[t]
\centering
\caption{
CNN configuration used for BCNN, SSR-CNN and MSR-CNN. Every convolution layer is followed by a leaky ReLU~\protect\cite{xu2015empirical} with 10$\%$ leakage. The architecture was optimized using 10$\%$ of training data as the validation set.
}
\label{table_arch_config}
\scalebox{0.8}{
\begin{tabular}{|c|c|c|c|c|c|c|}
\hline
Data-Set & \multicolumn{3}{c|}{MNIST and CIFAR-10/100} & \multicolumn{3}{c|}{Caltech-101 and ImageNet-2012} \\ \hline
Architecture & BCNN & SSR-CNN & MSR-CNN & BCNN & SSR-CNN & MSR-CNN \\ \hline
Input size & 28$\times$28$\times$1 & 28$\times$28$\times$1 & 28$\times$28$\times$1 & 224$\times$224$\times$3 & 224$\times$224$\times$3 & 224$\times$224$\times$3 \\ \hline
Subbands & - & 1-Layer & 1-Layer & - & 1-Layer & 1-Layer \\ \hline
CONV & 3$\times$3$\times$1$\times$64 & 3$\times$3$\times$4$\times$64 & 3$\times$3$\times$1$\times$(64/4)$\times$4 & 3$\times$3$\times$3$\times$64 & 3$\times$3$\times$12$\times$64 & 3$\times$3$\times$3$\times$(64/4)$\times$4 \\ \hline
CONV & 3$\times$3$\times$64$\times$128 & 3$\times$3$\times$64$\times$128 & \begin{tabular}[c]{@{}c@{}}3$\times$3$\times$(64/4)\\ $\times$(128/4)$\times$4\end{tabular} & 3$\times$3$\times$64$\times$64 & 3$\times$3$\times$64$\times$64 & 3$\times$3$\times$(64/4)$\times$(64/4)$\times$4 \\ \hline
CONV & 3$\times$3$\times$128$\times$256 & 3$\times$3$\times$128$\times$256 & \begin{tabular}[c]{@{}c@{}}3$\times$3$\times$(128/4) \\ $\times$(256/4)$\times$4\end{tabular} & 3$\times$3$\times$64$\times$64 & 3$\times$3$\times$64$\times$64 & 3$\times$3$\times$(64/4)$\times$(64/4)$\times$4 \\ \hline
CONV & - & - & - & 3$\times$3$\times$64$\times$64 & 3$\times$3$\times$64$\times$64 & 3$\times$3$\times$(64/4)$\times$(64/4)$\times$4 \\ \hline
CONV & - & - & - & 3$\times$3$\times$64$\times$64 & 3$\times$3$\times$64$\times$64 & 3$\times$3$\times$(64/4)$\times$(64/4)$\times$4 \\ \hline
POOL & 2-by-2 & 2-by-2 & 2-by-2 & 2-by-2 & 2-by-2 & 2-by-2 \\ \hline
CONV & 3$\times$3$\times$256$\times$512 & 3$\times$3$\times$256 $\times$512 & \begin{tabular}[c]{@{}c@{}}3$\times$3$\times$(256/4) \\ $\times$(512/4)$\times$4\end{tabular} & 3$\times$3$\times$64$\times$128 & 3$\times$3$\times$64$\times$128 & 3$\times$3$\times$(64/4) $\times$(128/4) $\times$4 \\ \hline
CONV & 3$\times$3$\times$512$\times$128 & 3$\times$3$\times$512$\times$128 & \begin{tabular}[c]{@{}c@{}}3$\times$3$\times$(512/4) \\ $\times$(128/4)$\times$4\end{tabular} & 3$\times$3$\times$128$\times$128 & 3$\times$3$\times$128$\times$128 & 3$\times$3$\times$(128/4) $\times$(128/4)$\times$4 \\ \hline
CONV & - & - & - & 3$\times$3$\times$128$\times$128 & 3$\times$3$\times$128$\times$128 & 3$\times$3$\times$(128/4)$\times$(128/4)$\times$4 \\ \hline
CONV & - & - & - & 3$\times$3$\times$128$\times$128 & 3$\times$3$\times$128$\times$128 & 3$\times$3$\times$(128/4)$\times$(128/4)$\times$4 \\ \hline
CONV & - & - & - & 3$\times$3$\times$128$\times$128 & 3$\times$3$\times$128$\times$128 & 3$\times$3$\times$(128/4)$\times$(128/4)$\times$4 \\ \hline
POOL & 2-by-2 & - & N/A & 2-by-2 & 2-by-2 & 2-by-2 \\ \hline
CONV & - & - & - & 3$\times$3$\times$64$\times$128 & 3$\times$3$\times$64$\times$128 & 3$\times$3$\times$(64/4) $\times$(128/4)$\times$4 \\ \hline
CONV & - & - & - & 3$\times$3$\times$128$\times$128 & 3$\times$3$\times$128$\times$128 & 3$\times$3$\times$(128/4) $\times$(128/4)$\times$4 \\ \hline
CONV & - & - & - & 3$\times$3$\times$128$\times$128 & 3$\times$3$\times$128$\times$128 & 3$\times$3$\times$(128/4) $\times$(128/4)$\times$4 \\ \hline
CONV & - & - & - & 3$\times$3$\times$128$\times$128 & 3$\times$3$\times$128$\times$128 & 3$\times$3$\times$(128/4) $\times$(128/4)$\times$4 \\ \hline
CONV & - & - & - & 3$\times$3$\times$128$\times$128 & 3$\times$3$\times$128$\times$128 & 3$\times$3$\times$(128/4) $\times$(128/4)$\times$4 \\ \hline
POOL & - & - & - & 2-by-2 & 2-by-2 & 2-by-2 \\ \hline
FC-1 & 7$\times$7$\times$128 $\times$4096 & 7$\times$7$\times$128 $\times$4096 & 7$\times$7$\times$128$\times$4096 & \begin{tabular}[c]{@{}c@{}}4$\times$4$\times$128$\times$4096 \\ OR 28$\times$28$\times$128$\times$4096\end{tabular} & \begin{tabular}[c]{@{}c@{}}4$\times$4$\times$128$\times$4096 \\ OR 14$\times$14$\times$128$\times$4096\end{tabular} & \begin{tabular}[c]{@{}c@{}}4$\times$4$\times$128$\times$4096 \\ OR 14$\times$14$\times$128$\times$4096\end{tabular} \\ \hline
DROPOUT & 50\% & 50\% & 50\% & 50\% & 50\% & 50\% \\ \hline
FC-2 & 4096$\times$1024 & 4096$\times$1024 & 4096$\times$1024 & 4096$\times$4096 & 4096$\times$1024 & 4096$\times$1024 \\ \hline
DROPOUT & 50\% & 50\% & 50\% & 50\% & 50\% & 50\% \\ \hline
FC-3 & 1024$\times$10 OR 1024$\times$101 & 1024$\times$10 OR 1024$\times$101 & 1024$\times$10 OR 1024$\times$101 & 4096$\times$101 OR 4096$\times$1000 & 4096$\times$101 OR 4096$\times$1000 & 4096$\times$101 OR 4096$\times$1000 \\ \hline
SOFTMA$\times$ & 1$\times$10 OR 1$\times$101 & 1$\times$10 OR 1$\times$101 & 1$\times$10 OR 1$\times$101 & 1$\times$101 OR 1$\times$1000 & 1$\times$101 OR 1$\times$1000 & 1$\times$101 OR 1$\times$1000 \\ \hline
\end{tabular}
}
\end{table*}

\section{Experimental Setup and Results}
\label{sec:Experimental_Setup_and_Results}
% write a line 
%To examine the performance of the considered architectures, we focus on an image classification problem. This is done solely for illustration purposes.
In this section, we investigate the performance of the proposed methods on image classification problems. We note that the proposed methods can work on any two-dimensional input signal such as image, radar, Lidar, sonar, etc.  

\subsection{Data Sets}
Our experiments use the MNIST, CIFAR-10/100, Caltech-101 and ImageNet-2012 datasets described in Table~\ref{data_set}. 
Images from both Caltech-101 and ImageNet-2012 datasets of varying sizes are resized to a common dimension of 256$\times$256$\times$3 using a Lanczos-3 kernel.
The only pre-processing done is data augmentation, where we randomly pick patches of size 224$\times$224$\times$3 from the four corners and the center of the image. 
The overlap between the images reflects translation in the images from patch to patch, thereby preventing data repetition in the training set. 
Further, to prevent data repetition, we add Gaussian blur~\cite{1057690} with a $3\times 3$ filter kernel and standard deviation randomly picked between 0.5 and 1.5.
%To the images, we add Gaussian noise with mean and variance equal to those of each subband. 
%This added subband channel distortion increases the total training image space tenfold. 
We found that without this data augmentation, the model is heavily overfitting.

\subsection{Models}
We compare the proposed architecture against two benchmark models: the SSR-CNN described in Section \ref{sec:ssrcnn}, and the BCNN model shown in Fig.~\ref{ssr-cnn}(b). 
The latter is a standard CNN that closely resembles AlexNet~\cite{NIPS2012_4824} and VGG-16~\cite{simonyan2015deep} and has ResNet-type~\cite{he2015deep} skip connections before each pooling layer.  
In contrast to BCNN, and similar to MSR-CNN, SSR-CNN also benefits from the exponential reduction of input data points per dimension, thereby substantially reducing the computational cost. 

The parameters that determine the learning capacity of a CNN include the number of convolution layers, filters per layer, activation function, arrangement of pooling layers, and the number of FC layers. 
Table~\ref{table_arch_config} shows the parameter values for the models, tuned heuristically due to the high computation cost of $K$-fold cross-validation. 
To study the effect of learning in the subspace-based CNN architectures, we keep most parameters constant across BCNN, MSR-CNN, and SSR-CNN. 
A network with five convolutional, two pooling, and three FC layers are used for MNIST.
A larger network of 15 convolutional, three pooling, and three FC layers are used for the CIFAR-10/100, Caltech-101, and ImageNet-2012 datasets.
In all three models, each convolutional layer uses small receptive field filters of size $3\times 3$ pixels~\cite{simonyan2015deep}.
A $2\times 2$ max pooling is chosen for the pooling layers. 
We use 50\% drop-out~\cite{NIPS2012_4824} at the first two FC layers to prevent significant overfitting, which helps reduce the difference in accuracy between training and test sets. 
We use no other regularization.
Finally, at the last FC layer, we use Softmax.   

For a fair comparison of MSR-CNN and SSR-CNN, we choose the total number of filters at each convolutional layer for both architectures to be equal, as seen in Table~\ref{table_arch_config}. 
In addition, the number of FC layers and their parameters are kept the same for both architectures.
%The number of filters at each layer of the 1-layer decomposed MSR-CNN is kept at 1/4th of BCNN per layer. 
The number of filters in each subband of the MSR-CNN is set to be equal to the total number of filters in the BCNN architecture divided by the number of subbands. 
For example, in a 1-layer decomposition structure, the number of filters in the first convolution layer of an MSR-CNN architecture, for each subband, is $1/4^{th}$ of the total number of filters in the first convolution layer of the BCNN architecture, 
%This is because there are four subbands in a 1-layer subband decomposition structure.
i.e., if the first layer in BCNN has 64 filters, then MSR-CNN has 16 filters per subband, as shown in Table~\ref{table_arch_config}.
For comparison purposes, for the WSD structure, we chose the Daubechies (D2) family of basis functions for DWT~\cite{57199}.

\subsection{Training}
We train our models using stochastic gradient descent with a mini-batch size of 64, batch-normalized, randomly picked images per mini-batch, momentum set to 0.9, and weight decay set to 0.0005 as indicated in~\cite{NIPS2012_4824}. 
The update equations at the $l$th iteration for the weights and biases of the convolutional filters at the $i$th layer and $k$th subband, $W_i^{k}(l)$, are given by:
\begin{flalign}
W_i^{k}(l+1) = W_i^{k}(l) + V_i^{k}(l+1)
\label{eq_7}
\end{flalign}
where
\begin{flalign}
V_i^{k}(l+1) &= 0.9\,V_i^{k}(l)-0.0005\,\epsilon\,W_i^{k}(l) -\, \epsilon\,  \left.\overline{\frac{\partial L}{\partial W_i^{k}}} \right| _{W_i^{k}(l)}
\label{eq_6}
\end{flalign}
Here $V_i^{k}(l)$ is the momentum, $\epsilon$ the learning rate, and $\left.\overline{\frac{\partial L}{\partial W}} \right| _{W_i^{k}(l)}$ is the average over the $l$th batch of the derivative of the objective function with respect to $W_i^k$, evaluated at $W_i^{k}(l)$. 
The learning rate is initialized to 0.01 and later reduced by 10 when the validation error stops improving. 
We found experimentally that over a wide range of training sets, using a learning rate of 0.1, the networks showed minimal learning or did not learn at all, while with a learning rate of 0.001, learning was slow. 
It seems the best choice of learning rate lies between 0.1 and 0.001, so, we used a learning rate of 0.01 without optimizing this hyper-parameter further. 
The weight update equations given in equations~(\ref{eq_7}) and~(\ref{eq_6}) also govern the weight update of the subband decomposition filters. 
We use a lower learning rate for the subband decomposition structure compared to the subband convolutional layers because we have noticed that otherwise, the overall network is unable to learn effectively.
We initialize the weights for both the subband structure and the CNN weights by drawing from a Gaussian distribution with a standard deviation of 0.01. All biases are initialized to 1. 

%Each model is trained on a GTX 1080Ti GPU with Nvidia CUDA \& cuDNN libraries. 
The following four parallel threads are run in the pipeline: (i) Read mini-batch from storage disk and re-size images, (ii) Compute data augmentation, (iii) Compute decomposition filters (2D-DWT/ASD/CASD) on CPU, and (iv) Transfer data to GPU, compute CNN on GPU, and read-back to system main memory. 
The GPU computation is the bottleneck, thereby resulting in almost free processing time for the rest of the parallel threads per batch. 
Pipeline fill-\&-flush are processed accordingly, with pipeline overhead being insignificant compared to total processing time.

During testing, while reading in the input image, we pick five patches from the input image, four from the four corners, and one from the center.
Predictions on these five patches are then averaged to obtain the final result of that single input image. 
Finally, we average over the entire test dataset to get the final test result.

\begin{figure}[h]
\includegraphics[scale=0.66]{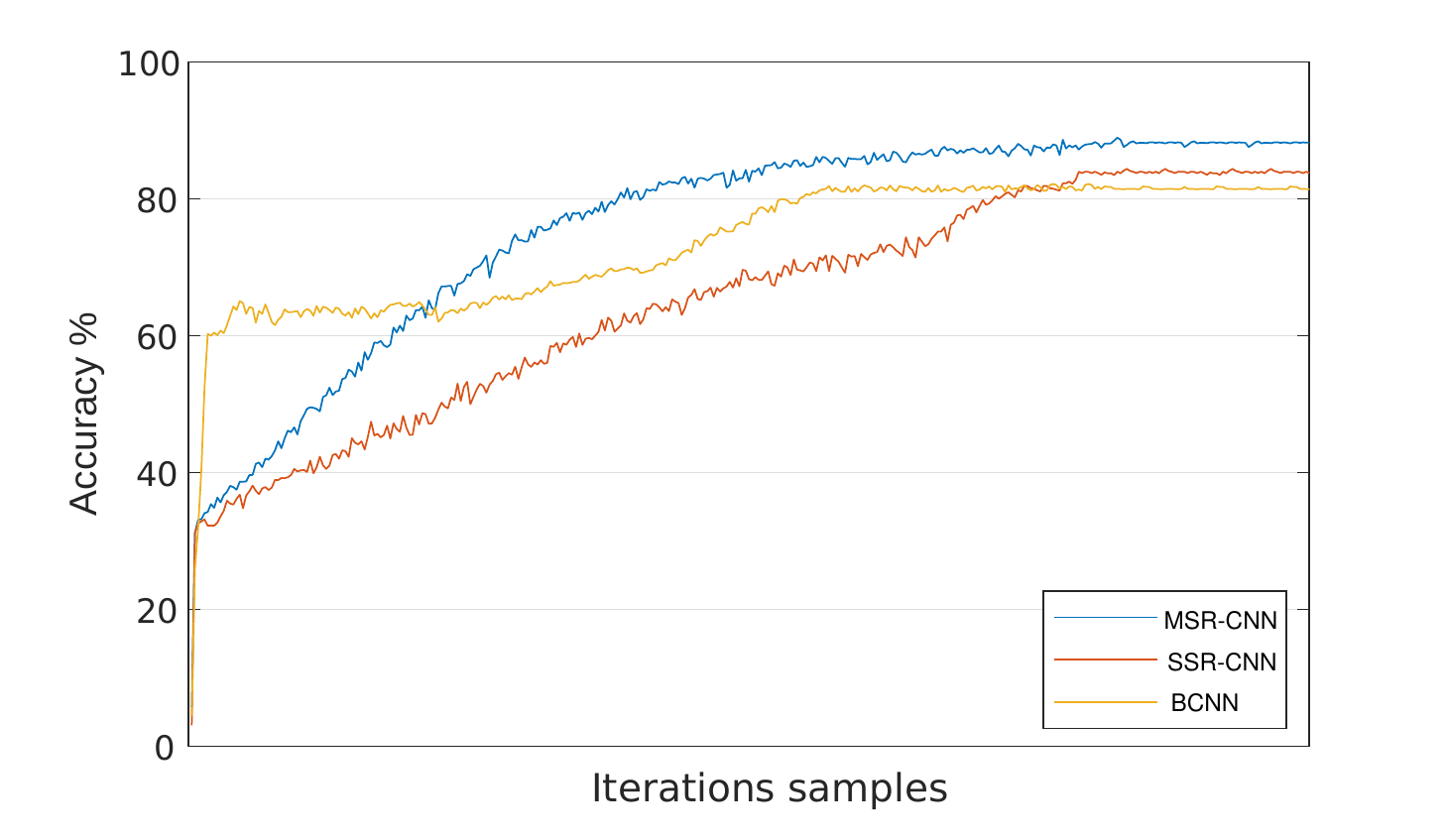}
\centering
\caption{Test accuracy comparison of BCNN and 1-Layer ASD structure of the SSR-CNN and the MSR-CNN architectures for the Caltech-101 dataset.}
\label{test_error_curve}
\end{figure}

\subsection{Results}
%\subsubsection{Frequency response of ASD filters}
\subsubsection{Qualitative analysis of subband decomposition}
In Fig.~\ref{fig_filt_resp}, we show examples of the 2D frequency responses of the first layer of the ASD filter structure (cf. Fig.~\ref{adaptive_filter}).
The top row shows the 2D frequency response of the filters in the first upper and lower paths that decompose the input signal into two subbands. 
The bottom row shows the 2D frequency response of the filters in the second set of upper and lower paths that completes the first layer decomposing the input into four subbands.
%Similarly, multiple layers can be decomposed.
Note that the first-row filter responses are low-pass (left) and high-pass (right), while the second row of filters consists of low-pass, wider range low-pass, band-pass, and high-pass
%high-pass, band-pass, wider range low-pass, and sharper low-pass 
filter frequency responses, from left to right, respectively. 
Interestingly, the learned filters in the proposed structure resemble subband decomposition filters, i.e., pairs of highpass and lowpass filters, typically used in signal processing. 

%Fig.~\ref{fig_filt_resp} shows the 2D-frequency response of the subband decomposed filter with adaptive subband (ASD) decomposition structure. 
Fig.~\ref{fig_in_img_DWT} shows the input image and the 4 subbands of a 1-layer 2D-DWT decomposition filter structure while Fig.~\ref{fig_in_img} shows the 4 subbands of a 1-layer ASD decomposition filter structure in MSR-CNN for the same input image. 
Fig.~\ref{output_layer_1} shows the output of the first convolutional layer of MSR-CNN with a 1-layer ASD structure. 
Though all subbands have a mix of low and high-frequency components, the subbands to the left have more low-frequency components while the subbands to the right have more high-frequency components as seen in the activation map. 
Fig.~\ref{output_layer_10} shows the output at the 10th convolutional layer. In contrast to Fig.~\ref{output_layer_1}, Fig.~\ref{output_layer_10} has lost the details in LL and UL subbands containing mainly low frequency or DC components, while the other two subbands still resemble the input and highlight specific contours of the input, indicating high-frequency components. The FC layer utilizes this collective information to partition the output hyperspace.

\subsubsection{Classification Accuracy}

In Table \ref{acc_percent} we present the top-1 accuracy  comparison across the BCNN, SSR-CNN, and MSR-CNN architectures, for both 1- and 2-layer ASD and CASD structures.
We have used a filter of order five for both the ASD and CASD filter structures. 
Generally, we notice a drop in performance going to a 2-layer decomposition compared to a 1-layer decomposition structure. This drop is a tradeoff between performance and computation cost.
We also observe that the ASD structure with the MSR-CNN performs the best across different datasets.

Fig.~\ref{test_error_curve} compares the test-set error curves of the BCNN, SSR-CNN, and MSR-CNN architectures. % with a 1-layer ASD structure for the Caltech-101 dataset.
Compared to both the BCNN and the SSR-CNN curves, we find that the accuracy of the MSR-CNN has a milder monotonically increasing trajectory before settling down to a higher accuracy value.
The BCNN increases quickly, then remains relatively stable, and finally shows an increase in accuracy before settling to a lower value compared to MSR-CNN and the SSR-CNN.
The SSR-CNN on the other hand makes an initial jump in accuracy like that of the MSR-CNN and thereafter improves in accuracy in an almost linear fashion until settling between the performance of the MSR-CNN and the BCNN.
%We find that the accuracy of both 1- and 2-layer ASD models converges closely and that MSR-CNN has a milder monotonically increasing graph before settling, compared to both BCNN and SSR-CNN architectures.
This indicates that, with a 1-layer ASD structure, the better regularization of the MSR-CNN architecture results in more effective training compared to the BCNN and the SSR-CNN architectures. 

On the ImageNet-2012 dataset, the MSR-CNN architecture with 1-layer ASD structure achieves top-5 and top-1 validation set accuracy %\footnote{See ~\cite{NIPS2012_4824} for the definition of top-5 and top-1 accuracy.}
 of 86.91\% and 69.45\%, respectively.
%This indicates better learning of the MSR-CNN architecture, which is majorly contributed by better regularization of the architecture.
%With a 1-layer ASD structure, the MSR-CNN was trained on the ImageNet-2012 dataset and achieves top-5 and top-1 validation set accuracy\footnote{See ~\cite{NIPS2012_4824} for the definition of top-5 and top-1 accuracy.} of 81.37\% and 63.7\%, respectively.
%AlexNet achieves an accuracy of 83\% and 62.5\%, respectively. 
Unlike most popular networks, we did not perform architectural hyper-parameter optimization due to the lack of heavy computational resources, leaving room for further improvements.

\begin{table}[t]
\caption{
Top-1 accuracy of BCNN, SSR-CNN, and MSR-CNN for layer-1 and layer-2 decomposition with the WSD, ASD, and CASD structures. CNN architecture parameters are detailed in Table~\ref{table_arch_config}.
}
\centering
\label{acc_percent}
\scalebox{0.85}{
\begin{tabular}{|cccc|}
\hline
\multicolumn{1}{|c|}{\textbf{Model}} & \multicolumn{1}{c|}{\textbf{Decomposition}} & \multicolumn{1}{c|}{\textbf{Dataset}} & \textbf{Accuracy \% (Top-1)} \\ \hline
\multicolumn{1}{|c|}{BCNN}           & \multicolumn{1}{c|}{-}                      & \multicolumn{1}{c|}{Caltech-101}      & 82.17                        \\ \hline
\multicolumn{1}{|c|}{SSR-CNN}        & \multicolumn{1}{c|}{1-layer (WSD)}          & \multicolumn{1}{c|}{Caltech-101}      & 83.89                        \\ \hline
\multicolumn{1}{|c|}{SSR-CNN}        & \multicolumn{1}{c|}{1-layer (CASD)}         & \multicolumn{1}{c|}{Caltech-101}      & 87.47                        \\ \hline
\multicolumn{1}{|c|}{SSR-CNN}        & \multicolumn{1}{c|}{1-layer (ASD)}          & \multicolumn{1}{c|}{Caltech-101}      & 89.05                        \\ \hline
\multicolumn{1}{|c|}{MSR-CNN}        & \multicolumn{1}{c|}{1-layer (WSD)}          & \multicolumn{1}{c|}{Caltech-101}      & 87.79                        \\ \hline
\multicolumn{1}{|c|}{MSR-CNN}        & \multicolumn{1}{c|}{1-layer (CASD)}         & \multicolumn{1}{c|}{Caltech-101}      & 88.13                        \\ \hline
\multicolumn{1}{|c|}{MSR-CNN}        & \multicolumn{1}{c|}{1-layer (ASD)}          & \multicolumn{1}{c|}{Caltech-101}      & 89.39                        \\ \hline
\multicolumn{1}{|c|}{SSR-CNN}        & \multicolumn{1}{c|}{2-layer (CASD)}         & \multicolumn{1}{c|}{Caltech-101}      & 79.67                        \\ \hline
\multicolumn{1}{|c|}{SSR-CNN}        & \multicolumn{1}{c|}{2-layer (ASD)}          & \multicolumn{1}{c|}{Caltech-101}      & 79.39                        \\ \hline
\multicolumn{1}{|c|}{MSR-CNN}        & \multicolumn{1}{c|}{2-layer (CASD)}         & \multicolumn{1}{c|}{Caltech-101}      & 80.94                        \\ \hline
\multicolumn{1}{|c|}{MSR-CNN}        & \multicolumn{1}{c|}{2-layer (ASD)}          & \multicolumn{1}{c|}{Caltech-101}      & 82.49                        \\ \hline
\multicolumn{4}{|c|}{}                                                                                                                                    \\ \hline
\multicolumn{1}{|c|}{BCNN}           & \multicolumn{1}{c|}{-}                      & \multicolumn{1}{c|}{ImageNet-2012}    & 63.21                        \\ \hline
\multicolumn{1}{|c|}{SSR-CNN}        & \multicolumn{1}{c|}{1-layer (CASD)}         & \multicolumn{1}{c|}{ImageNet-2012}    & 65.95                        \\ \hline
\multicolumn{1}{|c|}{SSR-CNN}        & \multicolumn{1}{c|}{1-layer (ASD)}          & \multicolumn{1}{c|}{ImageNet-2012}    & 66.84                        \\ \hline
\multicolumn{1}{|c|}{MSR-CNN}        & \multicolumn{1}{c|}{1-layer (CASD)}         & \multicolumn{1}{c|}{ImageNet-2012}    & 67.98                        \\ \hline
\multicolumn{1}{|c|}{MSR-CNN}        & \multicolumn{1}{c|}{1-layer (ASD)}          & \multicolumn{1}{c|}{ImageNet-2012}    & 69.45                        \\ \hline
\multicolumn{1}{|c|}{SSR-CNN}        & \multicolumn{1}{c|}{2-layer (CASD)}         & \multicolumn{1}{c|}{ImageNet-2012}    & 59.25                        \\ \hline
\multicolumn{1}{|c|}{SSR-CNN}        & \multicolumn{1}{c|}{2-layer (ASD)}          & \multicolumn{1}{c|}{ImageNet-2012}    & 59.67                        \\ \hline
\multicolumn{1}{|c|}{MSR-CNN}        & \multicolumn{1}{c|}{2-layer (CASD)}         & \multicolumn{1}{c|}{ImageNet-2012}    & 61.87                        \\ \hline
\multicolumn{1}{|c|}{MSR-CNN}        & \multicolumn{1}{c|}{2-layer (ASD)}          & \multicolumn{1}{c|}{ImageNet-2012}    & 62.73                        \\ \hline
\multicolumn{4}{|c|}{}                                                                                                                                    \\ \hline
\multicolumn{1}{|c|}{BCNN}           & \multicolumn{1}{c|}{-}                      & \multicolumn{1}{c|}{MNIST}            & 99.72                        \\ \hline
\multicolumn{1}{|c|}{SSR-CNN}        & \multicolumn{1}{c|}{1-layer (WSD)}          & \multicolumn{1}{c|}{MNIST}            & 99.76                        \\ \hline
\multicolumn{1}{|c|}{SSR-CNN}        & \multicolumn{1}{c|}{1-layer (CASD)}         & \multicolumn{1}{c|}{MNIST}            & 99.79                        \\ \hline
\multicolumn{1}{|c|}{SSR-CNN}        & \multicolumn{1}{c|}{1-layer (ASD)}          & \multicolumn{1}{c|}{MNIST}            & 99.83                        \\ \hline
\multicolumn{1}{|c|}{MSR-CNN}        & \multicolumn{1}{c|}{1-layer (WSD)}          & \multicolumn{1}{c|}{MNIST}            & 99.81                        \\ \hline
\multicolumn{1}{|c|}{MSR-CNN}        & \multicolumn{1}{c|}{1-layer (CASD)}         & \multicolumn{1}{c|}{MNIST}            & 99.83                        \\ \hline
\multicolumn{1}{|c|}{MSR-CNN}        & \multicolumn{1}{c|}{1-layer (ASD)}          & \multicolumn{1}{c|}{MNIST}            & 99.89                        \\ \hline
\multicolumn{4}{|c|}{}                                                                                                                                    \\ \hline
\multicolumn{1}{|c|}{BCNN}           & \multicolumn{1}{c|}{-}                      & \multicolumn{1}{c|}{CIFAR-10}         & 95.37                        \\ \hline
\multicolumn{1}{|c|}{SSR-CNN}        & \multicolumn{1}{c|}{1-layer (WSD)}          & \multicolumn{1}{c|}{CIFAR-10}         & 96.59                        \\ \hline
\multicolumn{1}{|c|}{SSR-CNN}        & \multicolumn{1}{c|}{1-layer (CASD)}         & \multicolumn{1}{c|}{CIFAR-10}         & 97.79                        \\ \hline
\multicolumn{1}{|c|}{SSR-CNN}        & \multicolumn{1}{c|}{1-layer (ASD)}          & \multicolumn{1}{c|}{CIFAR-10}         & 97.81                        \\ \hline
\multicolumn{1}{|c|}{MSR-CNN}        & \multicolumn{1}{c|}{1-layer (WSD)}          & \multicolumn{1}{c|}{CIFAR-10}         & 96.71                        \\ \hline
\multicolumn{1}{|c|}{MSR-CNN}        & \multicolumn{1}{c|}{1-layer (CASD)}         & \multicolumn{1}{c|}{CIFAR-10}         & 96.82                        \\ \hline
\multicolumn{1}{|c|}{MSR-CNN}        & \multicolumn{1}{c|}{1-layer (ASD)}          & \multicolumn{1}{c|}{CIFAR-10}         & 97.86                        \\ \hline
\multicolumn{4}{|c|}{}                                                                                                                                    \\ \hline
\multicolumn{1}{|c|}{BCNN}           & \multicolumn{1}{c|}{-}                      & \multicolumn{1}{c|}{CIFAR-100}        & 80.72                        \\ \hline
\multicolumn{1}{|c|}{SSR-CNN}        & \multicolumn{1}{c|}{1-layer (WSD)}          & \multicolumn{1}{c|}{CIFAR-100}        & 81.83                        \\ \hline
\multicolumn{1}{|c|}{SSR-CNN}        & \multicolumn{1}{c|}{1-layer (CASD)}         & \multicolumn{1}{c|}{CIFAR-100}        & 81.74                        \\ \hline
\multicolumn{1}{|c|}{SSR-CNN}        & \multicolumn{1}{c|}{1-layer (ASD)}          & \multicolumn{1}{c|}{CIFAR-100}        & 82.95                        \\ \hline
\multicolumn{1}{|c|}{MSR-CNN}        & \multicolumn{1}{c|}{1-layer (WSD)}          & \multicolumn{1}{c|}{CIFAR-100}        & 82.97                        \\ \hline
\multicolumn{1}{|c|}{MSR-CNN}        & \multicolumn{1}{c|}{1-layer (CASD)}         & \multicolumn{1}{c|}{CIFAR-100}        & 83.75                        \\ \hline
\multicolumn{1}{|c|}{MSR-CNN}        & \multicolumn{1}{c|}{1-layer (ASD)}          & \multicolumn{1}{c|}{CIFAR-100}        & 85.07                        \\ \hline
\end{tabular}
}
\end{table}

\subsubsection{Computational Cost}
In Table \ref{table_comp_reduction_inferance}, we compare the total computation cost for both inference and training operations, of MSR-CNN and SSR-CNN architectures with the ASD structure.
The reported numbers take into account every addition and multiplication operation performed by each network.
The total addition and multiplication operations involved in the convolution operations dominate the overall computational cost of the CNNs.
In the case of the back-propagation path, we take into account the number of operations {\em per iteration}.
In the case of the inference path, which is not iterative, we report the total number of operations.
%Relative to BCNN, the MSR-CNN architecture for inferencing with 1-layer ASD architecture obtains a 95.75\% reduction in total addition operations and a 95.47\% reduction in total multipliers operations. Similarly, a 2-layer decomposition has a decrease of 98.84\% and 98.76\% in the total addition and multiplication operations, respectively.
%The SSR-CNN architecture for training with 1-layer ASD decomposition obtains a 77.64\% reduction in total addition operations and a 76.11\% reduction in total multipliers operations. Similarly, a 2-layer decomposition has a decrease of 98.84\% and 98.76\% in the total addition and multiplication operations, respectively.
With a 1-layer ASD structure, the MSR-CNN achieves {\em nearly two orders of magnitude reduction} in total addition and multiplication operations relative to the BCNN.  
The total number of addition and multiplication operations reduces further going to a 2-layer subband decomposition structure.
When integrated over the period of training, it significantly reduces the total operations or the total energy spent.
The above significant computational cost reductions are directly attributed to the number of filters in each architecture. 
We note that the number of filters in the ASD filter is $2(4^{M}-1)$ filters, where $M$ is the number of layers. 
In comparison, the CASD filter structure requires $(4^{M}-1)$ filters. For example, a 3-layer ASD structure would have 126 such filters to compute, while the 3-layer CASD structure would only have 63 such filters to compute.

\begin{table}[t]
\caption{Computation reduction relative to BCNN at inference and training for 1-layer (1L) and 2-layer (2L) ASD SSR-CNN and MSR-CNN architectures with input dimension 224$\times$224$\times$3. 
The architectural configurations are described in Table \ref{table_arch_config}. %corresponding to the Caltech-101 and ImageNet-2012 data sets, 
%We take the BCNN architecture and transform it to its equivalent SSR-CNN and MSR-CNN architectures for comparison.
%"ADD-OPP" represents the total number of addition operations, and "MUL-OPP" represents the total number of multiplication operations performed by the CNNs, respectively.
}
\centering
\label{table_comp_reduction_inferance}
\scalebox{0.85}{
\begin{tabular}{ccc}
\hline
\multicolumn{1}{|l|}{Inference}             &\multicolumn{2}{|l|}{Total Reduction (\%) w.r.t BCNN} \\ \hline
\multicolumn{1}{|c|}{\multirow{2}{*}{}} & \multicolumn{1}{c|}{\multirow{2}{*}{Additions}} & \multicolumn{1}{c|}{\multirow{2}{*}{Multiplications}} \\
\multicolumn{1}{|c|}{}                      & \multicolumn{1}{c|}{}                       & \multicolumn{1}{c|}{}                     \\ \hline
\multicolumn{1}{|c|}{SSR-CNN (1L)}          & \multicolumn{1}{c|}{77.64}                  & \multicolumn{1}{c|}{76.11}                \\ \hline
\multicolumn{1}{|c|}{MSR-CNN (1L)}          & \multicolumn{1}{c|}{95.75}                  & \multicolumn{1}{c|}{95.47}                \\ \hline
\multicolumn{1}{|c|}{SSR-CNN (2L)}          & \multicolumn{1}{c|}{95.52}                  & \multicolumn{1}{c|}{95.22}                \\ \hline
\multicolumn{1}{|c|}{MSR-CNN (2L)}          & \multicolumn{1}{c|}{\textbf{98.84}}         & \multicolumn{1}{c|}{\textbf{98.76}}       \\ \hline                                                      
\multicolumn{1}{|l|}{Training}              &\multicolumn{2}{|l|}{Total Reduction (\%) w.r.t BCNN} \\ \hline
\multicolumn{1}{|c|}{\multirow{2}{*}{}} & \multicolumn{1}{c|}{\multirow{2}{*}{Additions}} & \multicolumn{1}{c|}{\multirow{2}{*}{Multiplications}} \\
\multicolumn{1}{|c|}{}                      & \multicolumn{1}{c|}{}                       & \multicolumn{1}{c|}{}                     \\ \hline
\multicolumn{1}{|c|}{SSR-CNN (1L)}          & \multicolumn{1}{c|}{72.84}                  & \multicolumn{1}{c|}{72.50}                \\ \hline
\multicolumn{1}{|c|}{MSR-CNN (1L)}          & \multicolumn{1}{c|}{75.93}                  & \multicolumn{1}{c|}{74.47}                \\ \hline
\multicolumn{1}{|c|}{SSR-CNN (2L)}          & \multicolumn{1}{c|}{81.11}                  & \multicolumn{1}{c|}{90.28}                \\ \hline
\multicolumn{1}{|c|}{MSR-CNN (2L)}          & \multicolumn{1}{c|}{\textbf{89.07}}         & \multicolumn{1}{c|}{\textbf{94.13}}       \\ \hline
\end{tabular}
}
\end{table}

\begin{table}[]
\centering
\caption{Comparison of the total time taken, number of MAC operations, number of parameters used, and classification accuracy of 1-\&2-layer MSR-CNN architecture with ASD structure with other well-established CNN models for the ImageNet-2012 dataset.
%The benchmark for calculating time in seconds was computed on a DELL PowerEdge R740xd server with 256 GB DDR4 2666MT/S, Dual CPU with 18 cores each Intel-Xeon Gold 5220, and an Nvidia GTX 1080TI GPU. 
The benchmark was computed on 1000 images with a batch size of 8 for all models.
}
\label{table_compute_benefit}
\scalebox{0.70}{
\begin{tabular}{|c|c|c|c|c|c|c|c|}
\hline
\textbf{Models} & \textbf{\begin{tabular}[c]{@{}c@{}}Time\\ (s)\end{tabular}} & \textbf{MACs}      & \textbf{\begin{tabular}[c]{@{}c@{}}Param. \\ (Millions)\end{tabular}} & \textbf{\begin{tabular}[c]{@{}c@{}}Param. \\ (MBytes)\end{tabular}} & \textbf{\begin{tabular}[c]{@{}c@{}}Accu.\\ Top-1\end{tabular}} & \textbf{\begin{tabular}[c]{@{}c@{}}Accu.\\ Top-5\end{tabular}} & \textbf{\begin{tabular}[c]{@{}c@{}}Diff\\ Top (5 - 1)\end{tabular}} \\ \hline
MobileNet V1~\cite{howard2017mobilenets}    	& 12.44                                                       & 569 (M)            & 4.24                                                                  & 2                                                                   & 70.9                                                           & 89.9                                                           & 19                                                                  \\ \hline
MobileNet V2~\cite{sandler2019mobilenetv2}    	& 14.63                                                       & 300 (M)            & 3.47                                                                  & 1.7                                                                 & 71.8                                                           & 91                                                             & 19.2                                                                \\ \hline
Google Net~\cite{szegedy2014going}      & 9.22                                                        & 741 (M)            & 6.99                                                                  & 3.3                                                                 & -                                                              & 92.1                                                           & -                                                                   \\ \hline
AlexNet~\cite{NIPS2012_4824}         						& 1.49                                                        & 724 (M)            & 60.95                                                                 & 29.1                                                                & 62.5                                                           & 83                                                             & 20.5                                                                \\ \hline
SqueezeNet~\cite{iandola2016squeezenet}      	& 8.48                                                        & 833 (M)            & \textbf{1.24}                                                         & \textbf{0.6}                                                        & 57.5                                                           & 80.3                                                           & 22.8                                                                \\ \hline
ResNet-50~\cite{he2015deep}       			& 6.47                                                        & 3.9 (B)            & 25.6                                                                  & 12.2                                                                & \textbf{75.2}                                                  & \textbf{93}                                                    & 17.8                                                                \\ \hline
VGG~\cite{simonyan2015deep}             		& 9.21                                                        & 15.5 (B)           & 138                                                                   & 65.8                                                                & 70.5                                                           & 91.2                                                           & 20.7                                                                \\ \hline
Inception-V1~\cite{DBLP:conf/cvpr/SzegedyLJSRAEVR15}    	& 2.45                                                        & 1.43 (B)           & 7                                                                     & 3.3                                                                 & 69.8                                                           & 89.3                                                           & 19.5                                                                \\ \hline
MSR-CNN (1L)  											& 1.52                                                        & \textbf{169.5 (M)} & 42.05                                                                 & 20.1                                                                & 69.45                                                          & 86.91                                                          & \textbf{15.68}                                                      \\ \hline
MSR-CNN (2L)  											& \textbf{0.79}                                               & \textbf{46.34 (M)} & 13.64                                                                 & 6.5                                                                 & 62.73                                                          & 79.15                                                          & 16.42                                                               \\ \hline
\end{tabular}
}
\end{table}

\begin{table}[t]
\caption{Top-1 inference accuracy with varying input quantization of the BCNN, SSR-CNN, and MSR-CNN architectures with 1-layer WSD, ASD, and CASD subband decomposition structures.}
\centering
\label{input_quant}
\scalebox{0.82}{
\begin{tabular}{|l|l|l|l|l|l|l|}
\hline
\textbf{Datasets}            & \textbf{Models} & \textbf{1-bit} & \textbf{2-bits} & \textbf{4-bits} & \textbf{6-bits} & \textbf{8-bits} \\ \hline
\multirow{7}{*}{MNIST}       & BCNN            & 97.42          & 97.53           & 97.6            & 97.69           & 99.72           \\ \cline{2-7} 
                             & SSR-CNN (WSD)   & 97.87          & 98.03           & 99.45           & 99.53           & 99.76           \\ \cline{2-7} 
                             & SSR-CNN (CASD)  & 96.23          & 98.03           & 99.47           & 99.58           & 99.79           \\ \cline{2-7} 
                             & SSR-CNN (ASD)   & 97.23          & 98.57           & 99.57           & 99.61           & 99.83           \\ \cline{2-7} 
                             & MSR-CNN (WSD)   & 99.5           & 99.6            & 99.63           & 99.64           & 99.81           \\ \cline{2-7} 
                             & MSR-CNN (CASD)  & 96.75          & 99.65           & 99.69           & 99.72           & 99.83           \\ \cline{2-7} 
                             & MSR-CNN (ASD)   & 99.5           & 99.69           & 99.71           & 99.76           & 99.89           \\ \hline
\multirow{7}{*}{CIFAR-10}    & BCNN            & 76.36          & 76.06           & 76.39           & 85.66           & 95.37           \\ \cline{2-7} 
                             & SSR-CNN (WSD)   & 76.66          & 76.38           & 79.2            & 89.56           & 96.59           \\ \cline{2-7} 
                             & SSR-CNN (CASD)  & 76.67          & 76.38           & 79.38           & 90.45           & 97.79           \\ \cline{2-7} 
                             & SSR-CNN (ASD)   & 76.66          & 76.38           & 87.45           & 93.49           & 97.81           \\ \cline{2-7} 
                             & MSR-CNN (WSD)   & 78.13          & 79.2            & 80.03           & 91.32           & 96.71           \\ \cline{2-7} 
                             & MSR-CNN (CASD)  & 79.04          & 80.56           & 84.76           & 92.25           & 96.82           \\ \cline{2-7} 
                             & MSR-CNN (ASD)   & 78.13          & 82.38           & 87.03           & 95.39           & 97.86           \\ \hline
\multirow{7}{*}{CIFAR-100}   & BCNN            & 54.89          & 59.66           & 68.8            & 73.85           & 80.72           \\ \cline{2-7} 
                             & SSR-CNN (WSD)   & 58.91          & 61.03           & 69.87           & 74.75           & 81.74           \\ \cline{2-7} 
                             & SSR-CNN (CASD)  & 59.79          & 63.78           & 70.98           & 76.76           & 82.83           \\ \cline{2-7} 
                             & SSR-CNN (ASD)   & 58.91          & 66.03           & 74.87           & 79.75           & 84.74           \\ \cline{2-7} 
                             & MSR-CNN (WSD)   & 60.14          & 64.79           & 69.15           & 74.07           & 82.97           \\ \cline{2-7} 
                             & MSR-CNN (CASD)  & 61.23          & 65.45           & 70.56           & 76.87           & 83.75           \\ \cline{2-7} 
                             & MSR-CNN (ASD)   & 60.14          & 66.79           & 72.15           & 77.07           & 85.07           \\ \hline
\multirow{7}{*}{CALTECH-101} & BCNN            & 69.87          & 71.96           & 76.2            & 79.31           & 82.17           \\ \cline{2-7} 
                             & SSR-CNN (WSD)   & 71.66          & 76.29           & 80.91           & 81.47           & 83.89           \\ \cline{2-7} 
                             & SSR-CNN (CASD)  & 69.89          & 76.14           & 81.04           & 82.67           & 84.07           \\ \cline{2-7} 
                             & SSR-CNN (ASD)   & 71.66          & 77.02           & 82.31           & 84.01           & 85.95           \\ \cline{2-7} 
                             & MSR-CNN (WSD)   & 71.17          & 74.94           & 77.61           & 83.16           & 86.93           \\ \cline{2-7} 
                             & MSR-CNN (CASD)  & 71.48          & 75.47           & 79.89           & 84.67           & 88.13           \\ \cline{2-7} 
                             & MSR-CNN (ASD)   & 72.17          & 75.94           & 79.94           & 84.74           & 89.39           \\ \hline
\end{tabular}
}
\end{table}

\begin{table}[t]
\caption{Top-1 inference accuracy with weight and bias quantization of the BCNN, SSR-CNN, and MSR-CNN architectures with 1-layer WSD, ASD, and CASD subband decomposition structures.}
\centering
\label{weight_quant}
\scalebox{0.82}{
\begin{tabular}{|l|l|l|l|l|}
\hline
\textbf{Datasets}            & \textbf{Models} & \textbf{8-bits} & \textbf{16-bits} & \textbf{32-bits} \\ \hline
\multirow{7}{*}{MNIST}       & BCNN            & 90.3            & 97.53            & 99.72            \\ \cline{2-5} 
                             & SSR-CNN (WSD)   & 90.91           & 97.9             & 99.76            \\ \cline{2-5} 
                             & SSR-CNN (CASD)  & 90.95           & 98.56            & 99.79            \\ \cline{2-5} 
                             & SSR-CNN (ASD)   & 90.91           & 99.01            & 99.83            \\ \cline{2-5} 
                             & MSR-CNN (WSD)   & 91.65           & 99.65            & 99.81            \\ \cline{2-5} 
                             & MSR-CNN (CASD)  & 91.95           & 99.66            & 99.83            \\ \cline{2-5} 
                             & MSR-CNN (ASD)   & 93.95           & 99.75            & 99.89            \\ \hline
\multirow{7}{*}{CIFAR-10}    & BCNN            & 60.85           & 76.13            & 95.37            \\ \cline{2-5} 
                             & SSR-CNN (WSD)   & 61.37           & 79.46            & 96.59            \\ \cline{2-5} 
                             & SSR-CNN (CASD)  & 63.45           & 81.94            & 97.79            \\ \cline{2-5} 
                             & SSR-CNN (ASD)   & 65.37           & 83.46            & 97.81            \\ \cline{2-5} 
                             & MSR-CNN (WSD)   & 63.54           & 79.84            & 96.71            \\ \cline{2-5} 
                             & MSR-CNN (CASD)  & 64.45           & 81.34            & 96.82            \\ \cline{2-5} 
                             & MSR-CNN (ASD)   & 63.54           & 79.84            & 97.86            \\ \hline
\multirow{7}{*}{CIFAR-100}   & BCNN            & 61.17           & 73.78            & 80.72            \\ \cline{2-5} 
                             & SSR-CNN (WSD)   & 61.13           & 61.13            & 81.83            \\ \cline{2-5} 
                             & SSR-CNN (CASD)  & 59.56           & 70.45            & 81.74            \\ \cline{2-5} 
                             & SSR-CNN (ASD)   & 61.13           & 71.13            & 82.95            \\ \cline{2-5} 
                             & MSR-CNN (WSD)   & 63.15           & 61.37            & 82.97            \\ \cline{2-5} 
                             & MSR-CNN (CASD)  & 61.65           & 70.02            & 83.75            \\ \cline{2-5} 
                             & MSR-CNN (ASD)   & 63.15           & 71.37            & 85.07            \\ \hline
\multirow{7}{*}{CALTECH-101} & BCNN            & 59.13            & 79.67            & 82.17            \\ \cline{2-5} 
                             & SSR-CNN (WSD)   & 75.4            & 80.01            & 83.89            \\ \cline{2-5} 
                             & SSR-CNN (CASD)  & 80.23           & 81.23            & 87.47            \\ \cline{2-5} 
                             & SSR-CNN (ASD)   & 75.4            & 80.01            & 89.05            \\ \cline{2-5} 
                             & MSR-CNN (WSD)   & 82.35           & 83.16            & 87.79            \\ \cline{2-5} 
                             & MSR-CNN (CASD)  & 81.23           & 84.23            & 88.13            \\ \cline{2-5} 
                             & MSR-CNN (ASD)   & 82.35           & 83.16            & 89.39            \\ \hline
\end{tabular}
}
\end{table}

\subsubsection{Comparisons to state-of-the-art}
In Table \ref{table_compute_benefit} we present a comprehensive comparison of well-established CNN architectures~\cite{howard2017mobilenets},\cite{sandler2019mobilenetv2},\cite{szegedy2014going},\cite{NIPS2012_4824},\cite{iandola2016squeezenet},\cite{he2015deep},\cite{simonyan2015deep}, and~\cite{DBLP:conf/cvpr/SzegedyLJSRAEVR15} against the best performer among the proposed CNN architectures. % with 1-layer and the 2-layer ASD structure with the MSR-CNN. 
The comparison is in terms of time taken, number of multiply-and-accumulate (MAC) units, number of total parameters, and CNN performance accuracy percentage for the ImageNet-2012 dataset.
The computation time was evaluated on a DELL PowerEdge R740xd server with 256 GB DDR4 2666MT/S, dual Intel-Xeon Gold 5220 CPUs with 18 cores each, and an Nvidia GTX 1080Ti GPU with Nvidia CUDA and cuDNN libraries. 
The results were obtained using 1000 images with a batch size of 8 images for all models.
As we can see, the MSR-CNN with the 1-layer ASD structure performs the best, with an accuracy of 86.91\% with 169.5M MAC operations and 42.05M parameters, which takes around 1.52 seconds to process 1000 images with a batch size of 8 images.

%Table \ref{table_compute_benefit} compares the total number of MAC operations and parameters used by state-of-the-art CNNs. 
Both the 1-layer and 2-layer subband decomposed MSR-CNNs give a best-in-class reduction in total MAC operations needed. 
The MSR-CNN with 1-layer decomposition takes 169.5M MAC operations while the 2-layer decomposition takes 46.34M MAC operations.
Other models range between 300M to 15.5B MAC operations.
On the number of parameters front, the MSR-CNN architecture performs fairly. 
The parameters of all compared CNNs require between 0.6 to 65.8 MBytes, with the 1-layer and 2-layer MSR-CNN architectures at 20.1 and 6.5 MBytes, respectively. 
In practice, the difference between 6.5 and 0.6 MBytes can be ignored, whereas a 10$\times$ reduction in total MAC operations can significantly improve computation time.
Finally, based on our experimental setup, the 1-layer, and 2-layer MSR-CNN take 1.52 secs and 0.79 sec respectively, % to compute 1000 images with a batch size of 8, 
while the other models range between 1.49 secs and 14.63 secs.

\subsubsection{Quantization Effect}
In Table \ref{input_quant}, we show the effect of different levels of input quantization for the BCNN, SSR-CNN, and MSR-CNN architectures with WSD, 1-layer CASD, and 1-layer ASD. 
We quantize the input at 1, 2, 4, 6, and 8 bits. 
%All internal nodes 
All internal computations are performed using the IEEE 32-bit floating-point precision.
The objective of this experiment is to analyze the effect of input quantization noise on CNN accuracy results over various CNN architectures, subband decomposition structures, quantization levels, and datasets. 
%This helps us analyze the effect of a single design parameter by isolating the contribution of the input quantization noise due to the particular design parameter on the CNN inference accuracy. 
%datasets, MNIST, CIFAR-10/100, and Caltech-101.
%In order to isolate the effect of input quantization noise on the overall CNN architecture as a result of the subband decomposition structure, we keep the CNN architecture constant and measure CNN performance for varying subband decomposition structures, keeping the decomposition layers to be constant and varying input quantization levels. 
We clearly observe that all three subband decomposition structures, WSD, CASD, and ASD are highly robust towards input quantization noise compared to the full-band BCNN architecture.
%We quantize the input to 1, 2, 4, 6, and 8 bits for this analysis.
%Table \ref{input_quant} summarizes the results. 
The results in Table \ref{input_quant} indicate that:
1) The subband decomposition structures provide robustness to input quantization error compared to full-band CNN architecture. 
2) The effect of input quantization on the subband decomposition architectures is not drastic, even when we go down to 1-bit quantization. 
This is not so surprising since even a human can fairly recognize an image well with 1-bit quantization.
For instance, grayscale newspaper printing used 1-bit greyscale with dithering~\cite{350813}.
Further, screens often use only 4-to-5 bits of color per channel natively and yet display good quality images, thanks to dithering~\cite{597270}.
Dithering fundamentally manipulates quantization noise by reshaping and spreading the noise unevenly across the spectrum. 
The subband decomposition process offers better input quantization even at 1-bit, compared to BCNN.
3) For applications requiring low transmission data rates or low storage and that intend to save power and computation resources, the flexibility offered by MSR-CNN with an adaptive subband decomposition front-end to quantize the input images to a greater extent without significantly affecting performance is advantageous from a system design consideration.

Weight-and-bias quantization plays a very important role in the implementation of CNNs in real-life systems~\cite{goyal2021fixedpoint}. 
Without weight and bias quantization, it would have been impractical to implement a CNN in an embedded system, where storage space and computation resources are limited. 
The ability to quantize at 8 bits compared to a 32-bit and 64-bit implementation is a matter of $1/4^{th}$ and $1/8^{th}$ savings in direct storage space, respectively.
Quantization of weight and bias also results in a lower memory footprint during the computation of the CNN model.
In Table \ref{weight_quant}, we investigate the effect of weight and bias quantization noise on inference accuracy for the BCNN, SSR-CNN, and MSR-CNN architectures with 1-layer WSD, CASD, and ASD for four different datasets, MNIST, CIFAR-10/100, and Caltech-101.
In this experiment, we consider different quantization levels such as 8-bit, 16-bit, and 32 bits.
All internal computations are performed using the IEEE 32-bit floating-point precision.
%Once again we vary a single design parameter such as subband decomposition structures (WSD, CASD, and ASD), CNN architectures (BCNN, SSR-CNN, and MSR-CNN), and quantization levels (8-, 16-, and 32-bit), keeping all other design parameters constant, per experiment.
%We see that the subband decomposition is robust to the quantization noise from the weights and biases.
%We isolate the quantization noise from weights and biases, by keeping the CNN architecture constant and varying the input decomposition structure, with a constant number of decomposition layers.
%We also quantize the weights and biases to 8, 16, and 32 bits for this analysis.
%Table \ref{weight_quant} summarizes the results.
Results from Table \ref{weight_quant} indicate that:
1) The subband decomposition structure provides robustness to weight and bias quantization error compared to full-band CNN architecture. 
2) The MSR-CNN with the ASD subband decomposition structure, in general, performs better than the other architectures.
3) The ability of MSR-CNN with an adaptive subband decomposition front-end to perform better with heavily quantized weights and biases makes it an attractive choice for applications that are highly sensitive to computation and storage resources. 

Both Tables VI and VII indicate that the MSR-CNN with the ASD decomposition structure performs the best. 
For applications with an additional need for reducing computation resources, the CASD structure may be a suitable compromise between computation cost and performance. As a final note, we would like to emphasize that the proposed architectures can be combined with sophisticated weight quantization methods, e.g., range batch normalization. Investigation of which method complements better the proposed architectures is beyond the scope of this work.

\section{Conclusion}
\label{sec:Conclusion}
We proposed two novel structures for learning the input signal decomposition into subbands: the ASD and the CASD structures. 
We incorporated these structures into a CNN architecture where each subband is individually processed via a separate CNN. 
The proposed structures integrate well with the end-to-end learning mechanism of a CNN. 
In the context of neural network-based image classification, we showed that the proposed CNN architecture achieves robust and near state-of-the-art performance compared to analyzing the entire spatial representation by a single CNN.
Importantly, this performance comes at less than $10\%$ of the computation cost of an equivalent full-band CNN. 
Three main factors contribute to such performance:
First, structural regularization is inherent to the proposed architecture.
Second, a given subband is protected from noise and other deformities, e.g., input quantization error and weight-and-bias quantization error, present in the rest of the subbands.
Third, the reduction of input dimensions of the convolution operation at each layer reduces the computational cost.
Ultimately, we would like to apply the efficient architecture to video sequence prediction, where computation cost grows exponentially due to the added dimension of time.

\bibliographystyle{IEEEtran}
\bibliography{my_bib}

\begin{thebibliography}{10}
\providecommand{\url}[1]{#1}
\csname url@samestyle\endcsname
\providecommand{\newblock}{\relax}
\providecommand{\bibinfo}[2]{#2}
\providecommand{\BIBentrySTDinterwordspacing}{\spaceskip=0pt\relax}
\providecommand{\BIBentryALTinterwordstretchfactor}{4}
\providecommand{\BIBentryALTinterwordspacing}{\spaceskip=\fontdimen2\font plus
\BIBentryALTinterwordstretchfactor\fontdimen3\font minus
  \fontdimen4\font\relax}
\providecommand{\BIBforeignlanguage}[2]{{%
\expandafter\ifx\csname l@#1\endcsname\relax
\typeout{** WARNING: IEEEtran.bst: No hyphenation pattern has been}%
\typeout{** loaded for the language `#1'. Using the pattern for}%
\typeout{** the default language instead.}%
\else
\language=\csname l@#1\endcsname
\fi
#2}}
\providecommand{\BIBdecl}{\relax}
\BIBdecl

\bibitem{NIPS2012_4824}
A.~Krizhevsky, I.~Sutskever, and G.~E. Hinton, ``Imagenet classification with
  deep convolutional neural networks,'' in \emph{Advances in Neural Information
  Processing Systems}, 2012, pp. 1097--1105.

\bibitem{long2015fully}
J.~Long, E.~Shelhamer, and T.~Darrell, ``Fully convolutional networks for
  semantic segmentation,'' 2015.[Online]. Available: arXiv:1411.4038.

\bibitem{vinyals2015tell}
O.~Vinyals, A.~Toshev, S.~Bengio, and D.~Erhan, ``Show and tell: A neural image
  caption generator,'' 2015.[Online]. Available: arXiv:1411.4555.

\bibitem{Toshev_2014}
\BIBentryALTinterwordspacing
A.~Toshev and C.~Szegedy, ``{DeepPose}: Human pose estimation via deep neural
  networks,'' in \emph{2014 {IEEE} Conference on Computer Vision and Pattern
  Recognition}.\hskip 1em plus 0.5em minus 0.4em\relax {IEEE}, jun 2014.
  [Online]. Available: \url{https://doi.org/10.1109%2Fcvpr.2014.214}
\BIBentrySTDinterwordspacing

\bibitem{Lecun98gradient-basedlearning}
Y.~Lecun, L.~Bottou, Y.~Bengio, and P.~Haffner, ``Gradient-based learning
  applied to document recognition,'' in \emph{Proceedings of the IEEE}, 1998,
  pp. 2278--2324.

\bibitem{zeiler2013visualizing}
M.~D. Zeiler and R.~Fergus, ``Visualizing and understanding convolutional
  networks,'' 2013.[Online]. Available: arXiv:1311.2901.

\bibitem{simonyan2015deep}
K.~Simonyan and A.~Zisserman, ``Very deep convolutional networks for
  large-scale image recognition,'' 2015.[Online]. Available: arXiv:1409.1556.

\bibitem{szegedy2014going}
C.~Szegedy, W.~Liu, Y.~Jia, P.~Sermanet, S.~Reed, D.~Anguelov, D.~Erhan,
  V.~Vanhoucke, and A.~Rabinovich, ``Going deeper with convolutions,''
  2014.[Online]. Available: arXiv:1409.4842.

\bibitem{he2015deep}
K.~He, X.~Zhang, S.~Ren, and J.~Sun, ``Deep residual learning for image
  recognition,'' 2015.[Online]. Available: arXiv:1512.03385.

\bibitem{jaderberg2016spatial}
M.~Jaderberg, K.~Simonyan, A.~Zisserman, and K.~Kavukcuoglu, ``Spatial
  transformer networks,'' 2016.[Online]. Available: arXiv:1506.02025.

\bibitem{DBLP:conf/cvpr/SzegedyLJSRAEVR15}
C.~Szegedy, W.~Liu, Y.~Jia, P.~Sermanet, S.~E. Reed, D.~Anguelov, D.~Erhan,
  V.~Vanhoucke, and A.~Rabinovich, ``Going deeper with convolutions,'' in
  \emph{{IEEE} Conf. Computer Vision and Pattern Recog., {CVPR}}, 2015, pp.
  1--9.

\bibitem{DBLP:journals/pami/HeZR015}
K.~He, X.~Zhang, S.~Ren, and J.~Sun, ``Spatial pyramid pooling in deep
  convolutional networks for visual recognition,'' \emph{{IEEE} Trans. Pattern
  Anal. Mach. Intell.}, vol.~37, no.~9, pp. 1904--1916, 2015.

\bibitem{Koch2015SiameseNN}
G.~Koch, R.~Zemel, and R.~Salakhutdinov, ``Siamese neural networks for one-shot
  image recognition,'' in \emph{32nd International Conference on Machine
  Learning}, 2015.

\bibitem{iandola2016squeezenet}
F.~N. Iandola, S.~Han, M.~W. Moskewicz, K.~Ashraf, W.~J. Dally, and K.~Keutzer,
  ``Squeezenet: Alexnet-level accuracy with 50x fewer parameters and <0.5mb
  model size,'' 2016.[Online]. Available: arXiv:1602.07360.

\bibitem{Parkhi2015DeepFR}
O.~M. Parkhi, A.~Vedaldi, and A.~Zisserman, ``Deep face recognition,'' in
  \emph{BMVC}, 2015.

\bibitem{cao2018vggface2}
Q.~Cao, L.~Shen, W.~Xie, O.~M. Parkhi, and A.~Zisserman, ``Vggface2: A dataset
  for recognising faces across pose and age,'' 2018.[Online]. Available:
  arXiv:1710.08092.

\bibitem{Schroff_2015}
F.~Schroff, D.~Kalenichenko, and J.~Philbin, ``{FaceNet}: A unified embedding
  for face recognition and clustering,'' in \emph{2015 {IEEE} Conference on
  Computer Vision and Pattern Recognition ({CVPR})}.\hskip 1em plus 0.5em minus
  0.4em\relax {IEEE}, jun 2015.

\bibitem{liu2022vision}
Y.~Liu, Y.-H. Wu, G.~Sun, L.~Zhang, A.~Chhatkuli, and L.~V. Gool, ``Vision
  transformers with hierarchical attention,'' 2022.[Online]. Available:
  arXiv:2106.03180.

\bibitem{NIPS2009_3848}
K.~Crammer, A.~Kulesza, and M.~Dredze, ``Adaptive regularization of weight
  vectors,'' in \emph{Advances in Neural Information Processing Systems
  22}.\hskip 1em plus 0.5em minus 0.4em\relax Curran Associates, Inc., 2009,
  pp. 414--422.

\bibitem{sun2015structure}
X.~Sun, ``Structure regularization for structured prediction: Theories and
  experiments,'' 2015.[Online]. Available: arXiv:1411.6243.

\bibitem{2018arXiv181000424T}
A.~{Tong}, D.~{van Dijk}, J.~S. {Stanley}, III, M.~{Amodio}, G.~{Wolf}, and
  S.~{Krishnaswamy}, ``{Graph Spectral Regularization for Neural Network
  Interpretability},'' \emph{ArXiv e-prints}, Sep. 2018.

\bibitem{6736905}
V.~N. Ekambaram, G.~Fanti, B.~Ayazifar, and K.~Ramchandran,
  ``Wavelet-regularized graph semi-supervised learning,'' in \emph{2013 IEEE
  Global Conference on Signal and Information Processing}, Dec 2013, pp.
  423--426.

\bibitem{doi:10.1117/12.208730}
S.-C.~B. Lo, H.~Li, J.-S. Lin, A.~Hasegawa, C.~Y. Wu, M.~T. Freedman, and S.~K.
  Mun, ``Artificial convolution neural network with wavelet kernels for disease
  pattern recognition,'' \emph{Proc.SPIE}, vol. 2434, pp. 1 -- 10, 1995.

\bibitem{Kang_2017}
E.~Kang, J.~Min, and J.~C. Ye, ``A deep convolutional neural network using
  directional wavelets for low-dose x-ray {CT} reconstruction,'' \emph{Medical
  Physics}, vol.~44, no.~10, pp. e360--e375, oct 2017.

\bibitem{7838150}
T.~Williams and R.~Li, ``Advanced image classification using wavelets and
  convolutional neural networks,'' in \emph{2016 15th IEEE International
  Conference on Machine Learning and Applications (ICMLA)}, 2016, pp. 233--239.

\bibitem{fujieda2018wavelet}
S.~Fujieda, K.~Takayama, and T.~Hachisuka, ``Wavelet convolutional neural
  networks,'' 2018.[Online]. Available: arXiv:1805.08620.

\bibitem{ulicny2022harmonic}
M.~Ulicny, V.~A. Krylov, and R.~Dahyot, ``Harmonic convolutional networks based
  on discrete cosine transform,'' 2022.[Online]. Available: arXiv:2001.06570.

\bibitem{Oyallon_2018}
E.~Oyallon, E.~Belilovsky, S.~Zagoruyko, and M.~Valko, ``Compressing the input
  for {CNNs} with the first-order scattering transform,'' in \emph{Computer
  Vision {\textendash} {ECCV} 2018}.\hskip 1em plus 0.5em minus 0.4em\relax
  Springer International Publishing, 2018, pp. 305--320.

\bibitem{bruna2012invariant}
J.~Bruna and S.~Mallat, ``Invariant scattering convolution networks,''
  2012.[Online]. Available: arXiv:1203.1513.

\bibitem{8804202}
P.~Sinha, I.~Psaromiligkos, and Z.~Zilic, ``A structurally regularized
  convolutional neural network for image classification using wavelet-based
  subband decomposition,'' in \emph{2019 IEEE International Conference on Image
  Processing (ICIP)}, 2019, pp. 649--653.

\bibitem{Wanhammar1999DSPIC}
L.~Wanhammar, \emph{\uppercase{DSP} \uppercase{I}ntegrated
  \uppercase{C}ircuits}.\hskip 1em plus 0.5em minus 0.4em\relax Elsevier, 1999.

\bibitem{5339431}
L.~D. Milić and J.~D. ćertić, ``Recursive digital filters and two-channel
  filter banks: Frequency-response masking approach,'' in \emph{2009 9th
  International Conference on Telecommunication in Modern Satellite, Cable, and
  Broadcasting Services}, 2009, pp. 177--184.

\bibitem{57199}
D.~I, ``The wavelet transform, time-frequency localization and signal
  analysis,'' \emph{IEEE Transactions on Information Theory}, vol.~36, no.~5,
  pp. 961--1005, 1990.

\bibitem{he2014convolutional}
K.~He and J.~Sun, ``Convolutional neural networks at constrained time cost,''
  2014.[Online]. Available: arXiv:1412.1710.

\bibitem{hoefler2021sparsity}
T.~Hoefler, D.~Alistarh, T.~Ben-Nun, N.~Dryden, and A.~Peste, ``Sparsity in
  deep learning: Pruning and growth for efficient inference and training in
  neural networks,'' 2021.

\bibitem{8771480}
S.~Kim, C.~Lee, H.~Park, J.~Wang, S.~Park, and C.~S. Park, ``Optimizations of
  scatter network for sparse cnn accelerators,'' in \emph{2019 IEEE
  International Conference on Artificial Intelligence Circuits and Systems
  (AICAS)}, 2019, pp. 256--257.

\bibitem{DBLP:journals/corr/ChangpinyoSZ17}
S.~Changpinyo, M.~Sandler, and A.~Zhmoginov, ``The power of sparsity in
  convolutional neural networks,'' \emph{CoRR}, vol. abs/1702.06257, 2017.

\bibitem{9034111}
Z.-G. Liu, P.~N. Whatmough, and M.~Mattina, ``Systolic tensor array: An
  efficient structured-sparse gemm accelerator for mobile cnn inference,''
  \emph{IEEE Computer Architecture Letters}, vol.~19, no.~1, pp. 34--37, 2020.

\bibitem{changpinyo2017power}
S.~Changpinyo, M.~Sandler, and A.~Zhmoginov, ``The power of sparsity in
  convolutional neural networks,'' 2017.[Online]. Available: arXiv:1702.06257.

\bibitem{DBLP:journals/corr/LiuWLC17}
C.~Liu, Y.~Wu, Y.~Lin, and S.~Chien, ``A kernel redundancy removing policy for
  convolutional neural network,'' \emph{CoRR}, vol. abs/1705.10748, 2017.

\bibitem{DBLP:journals/corr/LiKDSG16}
\BIBentryALTinterwordspacing
H.~Li, A.~Kadav, I.~Durdanovic, H.~Samet, and H.~P. Graf, ``Pruning filters for
  efficient convnets,'' \emph{CoRR}, vol. abs/1608.08710, 2016. [Online].
  Available: \url{http://arxiv.org/abs/1608.08710}
\BIBentrySTDinterwordspacing

\bibitem{DBLP:journals/corr/SzeCYE17}
V.~Sze, Y.~Chen, T.~Yang, and J.~S. Emer, ``Efficient processing of deep neural
  networks: {A} tutorial and survey,'' \emph{CoRR}, vol. abs/1703.09039, 2017.

\bibitem{1218191}
M.~Flierl and B.~Girod, ``Generalized b pictures and the draft h.264/avc
  video-compression standard,'' \emph{IEEE Transactions on Circuits and Systems
  for Video Technology}, vol.~13, no.~7, pp. 587--597, 2003.

\bibitem{1362510}
Y.~Xu and Y.~Zhou, ``H.264 video communication based refined error concealment
  schemes,'' \emph{IEEE Transactions on Consumer Electronics}, vol.~50, no.~4,
  pp. 1135--1141, 2004.

\bibitem{Lin:2016:FPQ:3045390.3045690}
D.~D. Lin, S.~S. Talathi, and V.~S. Annapureddy, ``Fixed point quantization of
  deep convolutional networks,'' in \emph{Intl. Conf. on Machine Learning},
  ser. ICML'16.\hskip 1em plus 0.5em minus 0.4em\relax JMLR.org, 2016, pp.
  2849--2858.

\bibitem{zhou2017incremental}
A.~Zhou, A.~Yao, Y.~Guo, L.~Xu, and Y.~Chen, ``Incremental network
  quantization: Towards lossless cnns with low-precision weights,''
  2017.[Online]. Available: arXiv:1702.03044.

\bibitem{zhou2017adaptive}
Y.~Zhou, S.-M. Moosavi-Dezfooli, N.-M. Cheung, and P.~Frossard, ``Adaptive
  quantization for deep neural network,'' 2017.[Online]. Available:
  arXiv:1712.01048.

\bibitem{lecun-mnisthandwrittendigit-2010}
\BIBentryALTinterwordspacing
Y.~LeCun and C.~Cortes, ``{MNIST} handwritten digit database,'' 2010. [Online].
  Available: \url{http://yann.lecun.com/exdb/mnist/}
\BIBentrySTDinterwordspacing

\bibitem{Krizhevsky09learningmultiple}
A.~Krizhevsky and G.~Hinton, ``Learning multiple layers of features from tiny
  images,'' University of Toronto, Toronto, Ontario, Tech. Rep.~0, 2009.

\bibitem{1384978}
L.~Fei-Fei, R.~Fergus, and P.~Perona, ``Learning generative visual models from
  few training examples: An incremental bayesian approach tested on 101 object
  categories,'' in \emph{2004 Conf. on Computer Vision and Pattern Recog.
  Workshop}, June 2004, pp. 178--178.

\bibitem{ILSVRC15}
O.~Russakovsky, J.~Deng, H.~Su, J.~Krause, S.~Satheesh, S.~Ma, Z.~Huang,
  A.~Karpathy, A.~Khosla, M.~Bernstein, A.~C. Berg, and L.~Fei-Fei, ``{ImageNet
  Large Scale Visual Recognition Challenge},'' \emph{International Journal of
  Computer Vision (IJCV)}, vol. 115, no.~3, pp. 211--252, 2015.

\bibitem{xu2015empirical}
B.~Xu, N.~Wang, T.~Chen, and M.~Li, ``Empirical evaluation of rectified
  activations in convolutional network,'' 2015.[Online]. Available:
  arXiv:1505.00853.

\bibitem{1057690}
R.~Graham, ``Snow removal--a noise-stripping process for picture signals,''
  \emph{IRE Transactions on Information Theory}, vol.~8, no.~2, pp. 129--144,
  1962.

\bibitem{howard2017mobilenets}
A.~G. Howard, M.~Zhu, B.~Chen, D.~Kalenichenko, W.~Wang, T.~Weyand,
  M.~Andreetto, and H.~Adam, ``Mobilenets: Efficient convolutional neural
  networks for mobile vision applications,'' 2017.[Online]. Available:
  arXiv:1704.04861.

\bibitem{sandler2019mobilenetv2}
M.~Sandler, A.~Howard, M.~Zhu, A.~Zhmoginov, and L.-C. Chen, ``Mobilenetv2:
  Inverted residuals and linear bottlenecks,'' 2019.[Online]. Available:
  arXiv:1801.04381.

\bibitem{350813}
T.~Pappas and D.~Neuhoff, ``Printer models and error diffusion,'' \emph{IEEE
  Transactions on Image Processing}, vol.~4, no.~1, pp. 66--80, 1995.

\bibitem{597270}
L.~Akarun, Y.~Yardunci, and A.~Cetin, ``Adaptive methods for dithering color
  images,'' \emph{IEEE Transactions on Image Processing}, vol.~6, no.~7, pp.
  950--955, 1997.

\bibitem{goyal2021fixedpoint}
R.~Goyal, J.~Vanschoren, V.~van Acht, and S.~Nijssen, ``Fixed-point
  quantization of convolutional neural networks for quantized inference on
  embedded platforms,'' 2021.[Online]. Available: arXiv:2102.02147.

\end{thebibliography}

\end{document}